\documentclass[aps,prl,twocolumn,showpacs,preprintnumbers,amsmath,amssymb,superscriptaddress]{revtex4-2}

\usepackage{graphicx}
\usepackage{dcolumn}
\usepackage{bm}
\usepackage{hyperref}
\hypersetup{
	colorlinks,
	citecolor=red,
	filecolor=black,
	linkcolor=blue,
	urlcolor=blue
}
\usepackage{lipsum}
\usepackage{subcaption}
\usepackage{subfloat}
\usepackage{caption}
\usepackage{natbib}

\newcommand{\nc}{\newcommand}
\nc{\be}{\begin{equation}}
\nc{\ee}{\end{equation}}
\nc{\bea}{\begin{eqnarray}}
\nc{\eea}{\end{eqnarray}}
\nc{\bean}{\begin{eqnarray*}}
\nc{\eean}{\end{eqnarray*}}
\nc{\mb}{\mbox}
\nc{\rnc}{\renewcommand}
\nc{\vk}{\mb{\bf k}}
\nc{\vp}{\mb{\bf p}}
\nc{\vn}{\mb{\bf n}}
\nc{\vq}{\mb{\bf q}}
\nc{\rr}{\mb{\bf r}}
\nc{\vz}{\hat {\mb{\bf z}}}
\nc{\vj}{\mb{\boldmath$j$}}
\nc{\vg}{\mb{\boldmath$g$}}
\nc{\x}{\mb{\boldmath$x$}}
\nc{\A}{\mb{\boldmath$A$}}
\nc{\va}{\mb{\boldmath$a$}}
\nc{\vs}{\mb{\boldmath$\sigma$}}
\nc{\vpi}{\mb{\boldmath$\pi$}}
\nc{\nab}{\nabla}
\nc{\X}{\sf x}
\nc{\kvec}{\mathbf{k}}
\nc{\pvec}{\mathbf{p}}
\nc{\qvec}{\mathbf{q}}
\nc{\vsp}{\vspace{0.05in}}
\nc{\vsps}{\vspace{--.-5in}}
\nc{\cfont}{\scriptsize}
\nc{\vspfig}{\vspace{-10pt}}
\nc{\vsigma}{\mb{\boldmath$\sigma$}}
\nc{\vder}{\mb{\boldmath$\nabla$}}
\nc{\vecq}{{\bf q}}
\nc{\veck}{{\bf k}}
\nc{\vecp}{{\bf p}}
\nc{\kk}{{\bd{k}}}
\nc{\pp}{{\bd{p}}}
\nc{\qq}{{\bd{q}}}

\nc{\argg}{\text{Arg}}
\nc{\bd}{\textbf}
\nc{\bds}{\boldsymbol}
\nc{\ham}{\hat{\mathcal{H}}}
\nc{\im}{\text{Im}}
\nc{\la}{\langle}
\nc{\ra}{\rangle}
\nc{\re}{\text{Re}}
\nc{\rn}[1]{%
	\textup{\uppercase\expandafter{\romannumeral#1}}%
}
\nc{\sgn}{\text{Sgn}}
\nc{\tit}{\textit}
\nc{\tr}{\text{tr}}
\nc{\changes}[1]{{\color{blue}{#1}}}

\def\be{\begin{eqnarray}}
\def\ee{\end{eqnarray}}

\begin{document}

\title{Pair Density Waves from Local Band Geometry}
\author{Guodong Jiang} 
\affiliation{Department of Physics, University of Nevada, Reno, Reno NV 89502, USA}
\author{Yafis Barlas}
\affiliation{Department of Physics, University of Nevada, Reno, Reno NV 89502, USA}

\begin{abstract}
A band-projection formalism is developed for calculating the superfluid weight in two-dimensional multi-orbital superconductors with an orbital-dependent pairing. It is discovered that, in this case, the band geometric superfluid stiffness tensor can be locally non-positive definite in some regions of the Brillouin zone. When these regions are large enough or include nodal singularities, the total superfluid weight becomes non-positive definite due to pairing fluctuations, resulting in the transition of a BCS state to a pair-density wave (PDW). This geometric BCS-PDW transition is studied in the context of two-orbital superconductors, and proof of the existence of a geometric BCS-PDW transition in a generic topological flat band is established.
\end{abstract}

\maketitle

{\em Introduction}.--- The superfluid weight of a superconductor is proportional to the density and inversely proportional to the effective mass~\cite{schrieffer1999theory}. Since it is the second-order expansion of the superconducting free energy in the center-of-mass momenta (CMM) of Cooper pairs, a positive definite superfluid weight ensures that zero CMM, the BCS state, minimizes the superconducting free energy.
This stability criterion is also satisfied by the recently discovered geometric superfluid weight in multi-orbital superconductors, which is proportional to the quantum metric of the band~\cite{peotta2015superfluidity,julku2016geometric,liang2017band,iskin2018exposing,torma2018quantum,wang2020quantum,julku2021quantum,
wang2021exact,rossi2021quantum,torma2022superconductivity,hu2019geometric,xie2020topology,Tian2023,hofmann2020superconductivity,
tovmasyan2016effective,herzog2022superfluid}. It is positive semidefinite for an orbital-independent order parameter~\cite{footnote1}, 
and dominates in flat band superconductors like twisted or strained two-dimensional (2D) crystals~\cite{bistritzer2011moire,SCLLstrainedgraphene,uchoa2017super,cao2018correlated,
cao2018unconventional,lu2019superconductors,sharpe2019emergent,yankowitz2019tuning,xu2018topological,
chen2019signatures,mora2019flatbands,julku2020superfluid,park2021tunable,wang2022hierarchy}. These rigid stability criteria seemingly preclude the existence of a pair density wave (PDW) state, which is an exotic superconductor whose electron pairs condense with a non-zero CMM~\cite{himeda2002,loder2010,lee2014,soto2014,fradkin2015colloquium,Agterberg2020,han2020,Liu2021}.

In this Letter, we show that the band geometric effect can stabilize a PDW state in flat bands with orbital-dependent pairing. Our analysis is based on the understanding of pairing instabilities that result from a negative definite superfluid weight $D_{s,\mu \nu}$ in multi-orbital superconductors. 
This criterion has recently been used to study the FFLO state in magnetic fields~\cite{kitamura2022quantum} and PDW states~\cite{chen2022pair}. However, the transition mechanism and its nature remain shrouded in mystery. Our theory, aided by developing a band projection formalism for the superfluid weight, allows us to identify a unique nodal mechanism that drives a BCS-PDW transition in the presence of orbital-dependent pairing. In the process, we establish an intimate connection between band topology and the BCS-PDW transition in superconductors.  

The nodal mechanism requires the presence of zeros in the quasiparticle spectrum. When these zeros coincide with the negative contributions to the superfluid weight ($D_{s,\mu \nu}$), they dominate over the positive contributions. We show that a topological two-orbital band always exhibits at least one nodal zero in the quasiparticle spectrum. Above a critical coupling, $D_{s,\mu \nu}$ becomes negative definite in the neighborhood of this nodal zero, driving a second-order BCS-PDW transition~\cite{footnote2}. To justify our claims with an explicit example, we study the BCS-PDW transitions in the flattened Bernevig-Hughes-Zhang (BHZ) model~\cite{bhz2006,qwz2006}. The PDW phase corresponds to the simultaneous presence of an attractive and repulsive channel, with the transition initiated by one channel turning repulsive. Below we sketch the analysis that leads to these results.

{\em Projected superfluid weight}.---We begin by describing our formalism for calculating the projected superfluid weight of multi-orbital superconductors with orbital-dependent pairing. Consider the 2D lattice Hamiltonian,
\be\label{eq:hlattice}
\ham=\sum_{ij,\alpha\beta,\sigma}h^{\sigma}_{ij,\alpha\beta}c^\dagger_{i\alpha\sigma}c_{j\beta\sigma}-\sum_{i,\alpha\beta}U_{\alpha\beta}c^\dagger_{i\alpha\uparrow}c^\dagger_{i\beta\downarrow}c_{i\beta\downarrow}c_{i\alpha\uparrow},
\ee
where $i,j$ denote for the Bravais lattice site, $\sigma=\uparrow,\downarrow$ for the spin, and $\alpha,\beta$ for orbitals or internal degrees of freedom other than spin. The generic single-particle Hamiltonian $h^{\sigma}_{ij,\alpha \beta}$ captures the hoppings between orbital $(i,\alpha)$ and $(j,\beta)$, where the spin-orbit coupling is ignored. The on-site pairing interaction $U_{\alpha\beta}$ is orbital-dependent and limited to singlet pairing. We assume that the Fermi energy $\mu$ lies within the $m^{th}$ band with spin-$\uparrow$ Bloch function $u_{m,\kk}$ and energy $\varepsilon_{m,\kk}$. Furthermore, to justify band projection, we assume that the interaction strength is of the order of bandwidth and much smaller than the band gap, $\delta\varepsilon_{m,\bd{k}}\lesssim|U_{\alpha \beta}|\ll E_{gap}$.

The superfluid weight tensor encodes the stability to the pairing fluctuation of $\bd{k}+\bd{q},\uparrow$ and $-\bd{k}+\bd{q},\downarrow$ electrons~\cite{footnote3}. 
For stability, it must be positive definite (PD). We perform the usual mean-field decoupling to arrive at the Bogoliubov-de-Genne (BdG) Hamiltonian $ \ham_{MF}(\hat{\Delta}) - \mu \hat{N}_e$, where $\hat{N}_e$ is the total electron number operator. The mean-field order parameter  $\hat{\Delta}_{\alpha\beta}=-U_{\alpha\beta}\la c_{i\beta\downarrow} c_{i\alpha\uparrow}\ra$ must be attained self-consistently (see the supplementary section). The projected grand potential for the $m^{th}$ band, $\Omega_{m}(\bd{q})$ can be expressed as,
\be\label{eq:grandproj}
\Omega_m(\bd{q})=-\frac{1}{\beta}\ln\tr \big\{ e^{-\beta \hat{\mathcal{P}}_m(\bd{q}) [\ham_{MF}(\hat{\Delta}_{\qq}) -\mu_\bd{q} \hat{N}_e ] \hat{\mathcal{P}}_m(\bd{q}) } \big\},
\ee
where $\hat{\mathcal{P}}_m(\bd{q})$ is a $\bd{q}$-dependent band-projection operator, and $\mu_\bd{q},\hat{\Delta}_{\bd{q}}$ are self-consistency functions coming from particle number constraint and the gap equation~\cite{peotta2015superfluidity,taylor2006pairing} (see the supplemental section). 

To calculate the $m^{th}$-band projected superfluid weight, $D^m_{s,\mu\nu}$, we assume that the translationally invariant BCS state is an extremum of the free energy. This is guaranteed when $\hat{\Delta}$ is Hermitian~\cite{footnote4}, 
giving $\tilde{\partial}_{q_\mu}|\Delta_{m,\bd{k}}(\bd{q})|^2|_{\bd{q}=0}=0$, where 
\be
\label{eq:projorderparameter}
\Delta_{m,\kk}(\qq) = \langle u_{m,\kk+\qq}| \hat{\Delta}_{\qq} | u_{m,\kk-\qq} \rangle,
\ee
is the band-projected gap function and symbol $\tilde{\partial}_{q_\mu}$ means the derivative does not act on $\mu_\bd{q}$ or $\hat{\Delta}_\bd{q}$, with $\mu = x,y$. Here, we focus on 2D systems at zero temperature; the finite-temperature case will be discussed elsewhere. The superfluid weight is computed from $D^m_{s,\mu\nu}=(1/N)\tilde{\partial}_{q_\mu}\tilde{\partial}_{q_\nu}\Omega_m(\bd{q})|_{\bd{q}=0}$,
where $\hbar$ and the unit cell area have been set to $1$, and $N$ is the number of unit cells. At $T=0$, it can be expressed as $ D^m_{s,\mu\nu} = D^{m,conv}_{s,\mu\nu} +D^{m,geo}_{s,\mu\nu}$,
\bea
\label{eq:dsconv}
D^{m,conv}_{s,\mu\nu} &=&\frac{1}{N}\sum_\bd{k}\bigg( 1- \frac{\xi_{m,\bd{k}}}{E_{m,\bd{k}}} \bigg) \partial_\mu \partial_\nu\xi_{m,\bd{k}}, \\
\label{eq:dsgeo}
D^{m,geo}_{s,\mu\nu} &=&  \frac{1}{N}\sum_\bd{k} \frac{G^m_{\mu\nu}(\bd{k})}{2E_{m,\bd{k}}},
\end{eqnarray}
where $G^m_{\mu\nu}(\bd{k})\equiv-\tilde{\partial}_{q_\mu}\tilde{\partial}_{q_\nu}|\Delta_{m,\bd{k}}(\bd{q})|^2|_{\bd{q}=0}$ is a gauge-invariant quantity that depends on both the Bloch function $u_{m,\bd{k}}$ and pairing matrix $\hat{\Delta}$, $\xi_{m,\bd{k}}=\varepsilon_{m,\bd{k}}-\mu$, and $E_{m,\bd{k}}=\sqrt{\xi^2_{m,\kk} + |\Delta_{m,\kk}(0)|^2}$. The conventional superfluid weight, $D^{m,conv}_{s,\mu\nu}$ is proportional to the band curvature and vanishes in the flat band limit. The generalized geometric superfluid weight, $D^{m, geo}_{s,\mu\nu}$ depends on the Bloch function $u_{m,\kk}$ through $G^m_{\mu \nu}$ which captures the band geometry.

When $\hat{\Delta}=\Delta_0 \hat{\mathcal{I}}$, we find $G^m_{\mu\nu}(\bd{k})=8|\Delta_0|^2g^m_{\mu\nu}(\bd{k})$, where $g^m_{\mu\nu}(\bd{k})$ is the quantum metric of the $m^{th}$ band, defined as the real part of quantum geometric tensor $R^m_{\mu\nu}(\bd{k})=\la\partial_\mu u_{m,\bd{k}}|(1-|u_{m,\bd{k}}\ra\la u_{m,\bd{k}}|)|\partial_\nu u_{m,\bd{k}}\ra$ \cite{provost1980riemannian}. This quantum-metric contribution was previously reported~\cite{peotta2015superfluidity,liang2017band} and is always PD. However, for a general pairing matrix $\hat{\Delta}$, tensor $G^m_{\mu\nu}(\bd{k})$ may be locally {\em non-positive definite} (NPD).

It is important to note that while $G^m_{\mu\nu}(\bd{k})$ is determined by the local band geometry and pairing matrix $\hat{\Delta}$, $D^{m,geo}_{s,\mu \nu}$, which determines the stability criteria, depends on both the band geometry and the quasiparticle spectrum $ E_{m,\bd{k}}$. This gives an intriguing admixture of band geometry and energetics. We find that the NPD behavior of $G^{m}_{\mu \nu}(\bd{k})$, even in part of the Brillouin zone (BZ),  can make $D^m_{s,\mu\nu}$ NPD, resulting in the pairing instability. 

{\em Instability toward a PDW state}.---To understand the physics associated with $D^{m,geo}_{s,\mu \nu}$, we analyze $G^m_{\mu\nu}(\bd{k})$ for a two-band system. We assume the order parameter matrix $\hat{\Delta}=\Delta_0\hat{\mathcal{I}}+\Delta_z\hat{\sigma}_z$, where $\Delta_0,\Delta_z$ are real and $\hat{\sigma}_z$ is the Pauli matrix on orbitals. This is equivalent to setting $U_{\alpha\beta}=\text{diag}(U_{11},U_{22})$ in the self-consistency equations. For a two-band Bloch Hamiltonian, $h^\uparrow(\bd{k})= \bd{h}(\kk) \cdot \boldsymbol{\sigma} $ with $\bd{h}(\kk) =(h_{x}(\kk),h_{y}(\kk),h_{z}(\kk))$, 
\bea\label{eq:gtensor}
G^{v(c)}_{\mu\nu}(\bd{k}) &=& 2 \Delta_0^2 \partial_\mu\hat{{\bd{h}}} \cdot \partial_\nu \hat{{\bd{h}}} \mp 2 \Delta_0 \Delta_z \partial_\mu\partial_\nu\hat{h}_z  \\ \nonumber
&-& 2\Delta_z^2(\partial_\mu \hat{h}_x\partial_\nu \hat{h}_x+\partial_\mu \hat{h}_y\partial_\nu \hat{h}_y + \hat{h} _z\partial_\mu\partial_\nu\hat{h}_z),
\end{eqnarray}
where $\hat{\bd{h}}=\bd{h}/|\bd{h}|$ and $v(c)$ denotes the valance (conduction) band. As noted above, the $\Delta_0^2$ term contains the quantum metric contribution, $g_{\mu\nu}(\bd{k})= (1/4) \partial_\mu\hat{{\bd{h}}} \cdot \partial_\nu \hat{{\bd{h}}}$. This PD term competes with the NPD contributions associated with $\Delta_0\Delta_z$ and $\Delta_z^2$ terms. It is easy to generalize Eq.~\ref{eq:gtensor} for a Hermitian $\hat{\Delta}$ (see supplemental section).

\begin{figure}[b]
\centering
\includegraphics[width=0.48\textwidth]{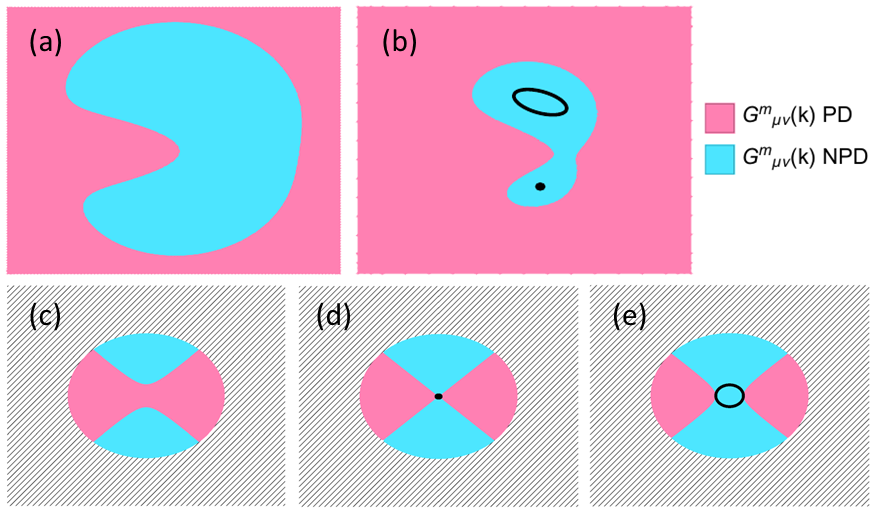}
\caption{(a)-(b) Schematic representations of two mechanisms leading to a NPD $D^{m,geo}_{s,\mu\nu}$. In (a), the $G^m_{\mu\nu}$ NPD region dominates over the PD region in the BZ, and in (b), The NPD $G^m_{\mu\nu}$ region encloses nodal points or circles. These singularities dominate even when the NPD area is small. (c)-(e) NPD $D^{m,geo}_{s,\mu\nu}$ driven by nodal circles for a generic topological flat band with (c) $\Delta_z/\Delta_0\lesssim 1$; (d) $\Delta_z/\Delta_0=1$; (e) $\Delta_z/\Delta_0\gtrsim 1$. Only the neighborhood of a nodal point is shown explicitly, while the rest of BZ is shaded.}
\label{fig:schematic}
\end{figure}

When the \bd{k}-space integral of the NPD contributions in Eq.~\ref{eq:dsgeo} dominates, $D^{m,geo}_{s,\mu \nu}$ becomes NPD (see Fig.~\ref{fig:schematic}) and the BCS state no longer minimizes the free energy. Since the free energy is defined over a compact domain of $\bd{q}$, it must attain an energy minimum at some $ \bd{Q} \neq 0 $, resulting in transitions to a PDW state. To estimate the transition point, consider a general class of two-band Hamiltonians: $h_z(\bd{k}) \to M$, with $M \gg|h_x|,|h_y|$. In a ``quasi-flat" band limit, \tit{defined} by condition $|\partial_\mu\partial_\nu(h_x^2+h_y^2)|\ll|\partial_\mu h_x\partial_\nu h_x|,|\partial_\mu h_y\partial_\nu h_y|$, $G_{\mu\nu}(\bd{k})$ becomes
\be
\label{eq:geobandmin}
G_{\mu\nu}(\bd{k})\approx \frac{2(\Delta_0^2 -\Delta_z^2)}{M^2}(\partial_\mu h_x\partial_\nu h_x+\partial_\mu h_y\partial_\nu h_y),
\ee
which is negative semidefinite for $ |\Delta_z/\Delta_0| \geq 1$, giving a NPD $D^{m,geo}_{s,\mu \nu}$. Therefore, $|\Delta_z/\Delta_0| \geq 1$ indicates the transition to PDW state.
As shown in Fig.~\ref{fig:schematic} (a), when the integrand of Eq.~\ref{eq:dsgeo} is a regular function over the BZ, this transition occurs when the NPD contributions (blue regions)  dominate the PD contributions (red regions). 

{\em PDW driven by nodal singularities}.---A more interesting scenario occurs in the presence of nodal zeroes (points or arcs) in the quasiparticle spectrum $E_{m,\kk}$. When these nodal zeroes are contained in the non-vanishing NPD regions of $G^m_{\mu \nu}$ in the BZ, as shown in Fig.~\ref{fig:schematic} (b), $D^{m,geo}_{\mu \nu}$ becomes singular and NPD. Analysis of nodal singularities shows that such a divergence requires $E_{m,\bd{k}} \propto k^{\alpha}$ with $\alpha \geq 2$ near nodal points and $\alpha \geq 1$ near nodal arcs, where $k$ is the distance from $\bd{k}$ to the nodal point or arc. In this case, the conventional superfluid weight, which is always regular, can be ignored, indicating that the BCS-PDW transition will occur as long as the NPD regions of $G^m_{\mu \nu}$ enclose these nodal zeroes. 

The valence-band-projected order parameter for $\hat{\Delta}=\Delta_0\hat{\mathcal{I}}+\Delta_i\hat{\sigma}_i $ can be calculated from Eq.~\ref{eq:projorderparameter}, giving $\Delta_{v,\bd{k}}=\Delta_{0} - \Delta_{i} \hat{h}_i(\bd{k})$, where $i=x,y,z$. In a flat band, when $\Delta_i = \Delta_0$, an isolated nodal point appears at $\bd{k}_0$ if $\hat{h}_i(\bd{k}_0)=1$. If the band is also topological, the map of $\bd{k}\mapsto\hat{\bd{h}}(\bd{k})$ wraps the entire Bloch sphere, {\em hence a nodal point singularity is always present at $\Delta_i =\pm\Delta_0$, for any hybridization direction $i$}. Near the nodal point $\bd{k}_0$, $G^v_{xx}(\bd{k})$ for the case $i=z$ can be expanded as,
\be
\label{eq:geonodal}
G^v_{xx}(\bd{k}) \approx - v^2\Delta_0^2\big[ v^2(p_y^2 - p_x^2) + 2\eta(1 + v^2 (p_y^2 + p_x^2))\big]
\ee
to the leading order in $\eta$, where $\bd{p} = \bd{k} - \bd{k}_0$, $\eta = g-1$ with $ g=\Delta_z/\Delta_0$, and $v > 0$ is an expansion coefficient (see supplemental section). At $g =1$, the nodal point is regular since $G^v_{\mu\nu}(\bd{k}) \to 0$ (Fig.~\ref{fig:schematic} (d)). However, when $g \gtrsim1$, the nodal point expands to a circle, which is \tit{contained} in the NPD $G^v_{\mu\nu}$ regions (Fig.~\ref{fig:schematic} (e)). This negative singular contribution dominates in the $\bd{k}$-space integral of Eq.~\ref{eq:dsgeo}, resulting in a negative divergent $D^v_s$ for $g > 1$. Since this behavior is determined from topological considerations, it applies to a generic topological two-band model with a Hermitian pairing matrix $\hat{\Delta}$.

{\em BCS-PDW transition in the BHZ model}.---To explore the BCS-PDW transitions in detail, we study the BCS instability in the flattened BHZ model, for a $\hat{\Delta}=\Delta_0\hat{\mathcal{I}}+\Delta_z\hat{\sigma}_z$. To keep the focus on the geometric term, we flatten the model by taking $h^\uparrow(\bd{k})=\epsilon_0\hat{\bd{h}}\cdot\bds{\sigma}$, where $\hat{\bd{h}}$ is the unit vector proportional to $(\sin k_x,\sin k_y,m_0+\cos k_x+\cos k_y)$. The band gap $2\epsilon_0$ is assumed to be larger than the interaction energy $U_{\alpha\beta}$, and $m_0$ is the scaled BHZ mass. Spin-orbit coupling is ignored, so the two spin components are decoupled and related by time-reversal symmetry (TRS). The BHZ model has a rich phase diagram with four phases---two topological phases $|m_0| < 2$ with Chern number $C =-sgn(m_0)$, and two trivial phases $|m_0| > 2$. This flattened Hamiltonian is invalid at the ``gapless" phase boundary $m_0=0,\pm2$, where the model undergoes a topological phase transition. For our calculations, we project to the valence band and take $\mu$ to be around $-\epsilon_0$.

In the flattened BHZ model, nodal zeroes occur when the projected band gap vanishes. When $\Delta_z/\Delta_0>0$, this happens for $m_0 > -2$ phases, where the BCS theory predicts nodal superconductors; for $m_0 < -2$, the superconductor is fully gapped. The three phases $m_0 >2$, $0<m_0< 2 $ and $-2<m_0<0$, have $4$, $3$ and $1$ nodal zeroes, respectively. To cure the divergence, we calculate $D^v_{s,\mu\nu}$ with both a small broadening $\xi_{v,\bd{k}}=\epsilon$ and weak dispersion $\xi_{v,\bd{k}}=\delta\xi_\bd{k}$ in the quasiparticle energy separately, hereafter referred to as $\epsilon$-broadened and dispersion-broadened.

\begin{figure}[b]
\centering
\includegraphics[width=0.48\textwidth]{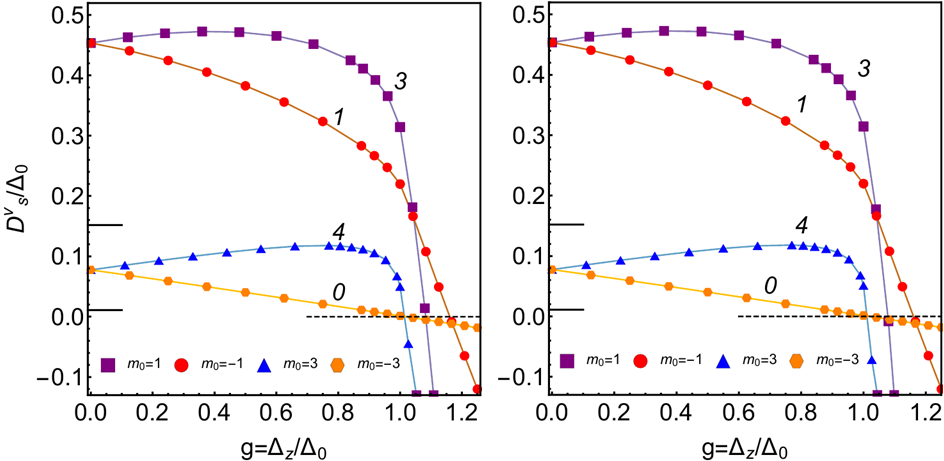}
\caption{Superfluid weight $D^v_{s}$ vs. $\Delta_z/\Delta_0 (>0)$ for $\Delta_0\hat{\mathcal{I}}+\Delta_z\hat{\sigma}_z$ type pairing. Four distinct BHZ phases $m_0=\pm1,\pm3$ are shown, with their number of nodal circles indicated by the numerical values. The horizontal lines indicate the topological lower bound for the two topological and trivial phases. (a) With broadening $\epsilon=0.01\Delta_0$, $E_{v,\bd{k}}=\sqrt{\epsilon^2+|\Delta_{v,\bd{k}}(0)|^2}$. (b) With weak dispersion of bandwidth $\delta\xi_\bd{k}\sim0.01\Delta_0$, $E_{v,\bd{k}}=\sqrt{\delta\xi_\bd{k}^2+|\Delta_{v,\bd{k}}(0)|^2}$. In this case, the conventional superfluid weight Eq.~\ref{eq:dsconv} is also included.}
\label{fig:zdxx}
\end{figure}

Fig.~\ref{fig:zdxx} (a) and (b) show the superfluid weight as a function of $g=\Delta_z/\Delta_0$ in various BHZ phases for the $\epsilon$-broadened and dispersion-broadened cases, respectively. Since $\sigma_z$ pairing term preserves the $C_4$ rotation symmetry, $D^v_{s,xx}=D^v_{s,yy}= D^v_s$. At some critical point $g=g_c$, $D^v_s$ becomes negative, indicating a transition to the PDW state. The four values of $m_0=\pm1,\pm3$ represent the distinct BHZ phases. The horizontal black lines indicate the lower-bound for $D^v_{s}$ at $\Delta_z=0$, where the quantum metric is lower-bounded by the absolute value of Berry curvature \cite{peotta2015superfluidity}. For this reason, the topological phases $m_0=\pm1$ have larger $D^v_{s}$ than the trivial phases $m_0=\pm3$ at small $g$ values.

In Fig.~\ref{fig:zdxx} (a) and (b), the superfluid weight for the $m_0=3,\pm1$ phases is marked by a sharp downturn of $D^v_{s}$ near $g=1$. This behavior can be attributed to nodal circles of quasiparticle energy $E_{v,\bd{k}} $ (Fig.~\ref{fig:schematic}(c)-(e)). We find no qualitative differences between the curves of the $\epsilon$-broadened and dispersion-broadened cases. More nodal circles results in a steeper slope of $D^v_s$ at $g>1$. In contrast, the $m_0=-3$ phase has no nodal zeros for $g\geq1$. Hence $D^v_{s}$ decreases steadily. 

\begin{figure}[b]
\centering
\includegraphics[width=0.45\textwidth]{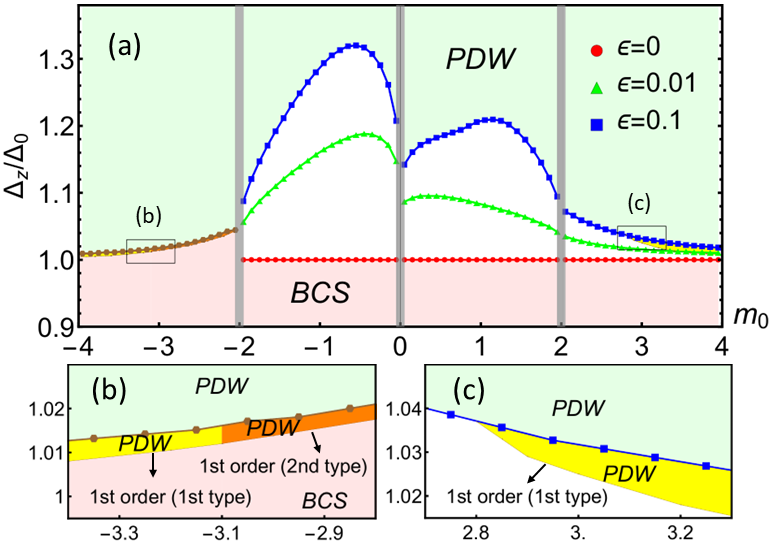}
\caption{Phase diagram of the flattened BHZ model with $\Delta_0\hat{\mathcal{I}}+\Delta_z\hat{\sigma}_z$ pairing. (a) $D^v_s$ instability curves for broadening $\epsilon=0$, $0.01$ and $0.1$ are shown in red, green and blue, respectively; they coincide in the $m_0<-2$ phase so are shown in brown. The $\epsilon=0$ data for $m_0>-2$ phases come from nodal singularity analysis. For second-order transitions, these curves are identical to the BCS-PDW phase boundary. The BCS region $0<\Delta_z/\Delta_0<0.9$ is snipped. (b)-(c) Magnified plots showing the first-order phase transition details with the same axis labels as (a). $\epsilon=0.01$ is similar to $\epsilon=0.1$, so only $\epsilon=0.1$ data is shown. (b) Two types of first-order transitions, at $-\infty<m_0<-2$, are separated at $m_0\sim-3.1$. (c) The first type of first-order transition at $2.85<m_0<+\infty$.}
\label{fig:zdigram}
\end{figure}

Fig.~\ref{fig:zdigram} shows the BCS-PDW phase boundary determined from the instability condition $D^v_{s}=0$ as a function of $m_0$. $D^v_s$ becomes negative at $g_c \gtrsim1$, for all $m_0$ values, especially $g_c \rightarrow1$ as $m_0\rightarrow\pm\infty$. This asymptotic behavior can be confirmed by an analysis similar to the previous discussion of a ``quasi-flat" band. Fig.~\ref{fig:zdigram} also shows the dependence of the phase boundary on $\epsilon$. $\epsilon$-broadening smears the nodal singularities of all three BHZ phases at $m_0>-2$, requiring a larger $g$ value for the PDW transition. The $m_0<-2$ phase has no nodal zeroes, so the phase boundary is unaffected. A careful examination of the free energy as a function of CMM $\bd{q}$ shows that the $D^v_s$ instability curves coincide with the BCS-PDW phase boundary only when the transition is second-order. There are also two types of weak first-order transitions located at $m_0<-2$ and $m_0>2.85$, which are slightly below the instability curve. These findings are summarized in Fig~\ref{fig:zdigram} (b) and (c).

{\em Pair density wave}.---To understand the nature of the BCS-PDW phase transition and the properties of the PDW state, we calculate the free energy per unit cell, $F_v(\bd{q})/N$ as a function of the CMM $\bd{q}$. The free energy is $F_v(\bd{q})=\Omega_v(\bd{q})+\mu_\bd{q}N_{e,v}$, with $N_{e,v}$ the total electron number in the valence band and $\Omega_v(\bd{q})$ evaluated from Eq.~\ref{eq:grandproj} at $T=0$ (see the supplemental section for details). $F_v(\bd{q})$ has the periodicity of time-reversal-invariant momentum (TRIM), which TRS imposes. Therefore we only need to focus on the region enclosed by $\Gamma$ $(0,0)$, $X$ $(\pi,0)$, $M$ $(\pi,\pi)$ and $X'$ $(0,\pi)$ in $\bd{q}$ space. As $D^v_s$ turns negative, these four high symmetry points are no longer local minima of $F_v(\bd{q})$, and a new minimum at $\bd{Q}$ corresponding to the CMM of PDW state emerges in the BZ.

\begin{figure}[b]
\centering
\includegraphics[width=0.49\textwidth]{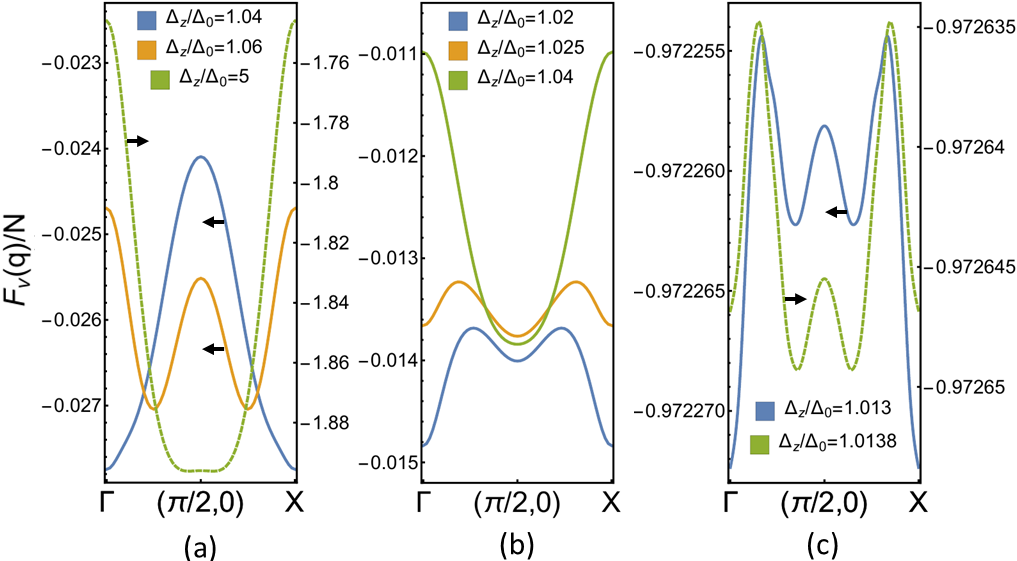}
\caption{$F_v(\bd{q})/N$ (in unit of $\Delta_0$) along $\Gamma X$ in $\bd{q}$ space, for three types of BCS-PDW transitions. (a) The second-order transition at $\epsilon=0.1,m_0=2.5$, $Q\neq \pi/2$; (b) first-order transition (first type) at $\epsilon=0.1,m_0=3$, $Q=\pi/2$; (c) first-order transition (second type) at $\epsilon=0.1,m_0=-3$, $Q\neq\pi/2$.}
\label{fig:symline}
\end{figure}

In general, the CMM $\bd{Q}$ is a function of model parameters and depends on the transition type. In continuous models, the transition type may depend on the higher-order derivatives of free energy \cite{casalbuoni2004inhomogeneous,kitamura2022quantum}; here in a lattice, it has to be determined from the calculation of $F_v(\bd{q})$ in all of $\bd{q}$-space. For our model, $C_4$ symmetry gives that $\bd{Q}$ modulo TRIM can be one of the four vectors ${\bf Q}_i=\{(\pm Q,0),(0,\pm Q)\}$, giving a set of biaxial PDW orders $ \{\Delta_{\bd{Q}_1},\Delta_{\bd{Q}_2},\Delta_{\bd{Q}_3},\Delta_{\bd{Q}_4}\} $. In Fig.~\ref{fig:symline}, we show three prototypes of BCS-PDW transitions, at parameter $m_0=2.5, 3$ and $-3$, respectively (c.f. Fig.~\ref{fig:zdigram}). The PDW phase corresponds to $U_{11} U_{22} < 0$, and the BCS-PDW transition coincides with one channel turning repulsive. The full phase diagram in the $(U_{11},U_{22})$ parameter space is provided in the supplemental section.

Second-order transitions (Fig.~\ref{fig:symline}(a)) are found in the middle region $-2<m_0<2.85$ of the phase diagram, with a maximum of free energy appearing at $(\pi/2,0)$. The transition occurs precisely when $D^v_{s}$ becomes negative, creating two minima $(Q,0),(\pi-Q,0)$ on $\Gamma X$, resulting in an incommensurate PDW. Notice $(\pi-Q,0)$ is equivalent to $(-Q,0)$ by shifting a TRIM. As $\Delta_z/\Delta_0$ becomes large, $Q$ tends to $\pi/2$, and the two minima merge into one, possibly converging to a commensurate PDW.

As shown in Fig.~\ref{fig:zdigram} (b) and (c), the BCS-PDW transitions for the flattened BHZ model can also be first-order. There are two types of first-order transitions, depending on the presence of minima or maxima at $(\pi/2,0)$. If it is a minimum (Fig.~\ref{fig:symline}(b)), then $Q=\pi/2$ is a constant and competes with the $\bd{q}=0$ BCS state. This first type of first-order transitions is found in the phase diagram's atomic regions $m_0<-3.1$ or $>2.85$, and results in a commensurate PDW.  Otherwise, if it is a maximum (Fig.~\ref{fig:symline}(c)), then there are two minima $(Q,0),(\pi-Q,0)$ competing with the $\bd{q}=0$ state, resulting in an incommensurate PDW. This second type can be viewed as a hybrid of the previous two types of transitions and is located in the intermediate region $-3.1<m_0<-2$. As $\Delta_z/\Delta_0$ becomes large, $Q\rightarrow\pi/2$ is also observed.

{\em Discussion}.--- The geometric instabilities discussed here are relevant to various weakly dispersive or flat band multi-orbital superconductors, for example, alternating twisted graphene-based superconductors which exhibit superconducting diode effect~\cite{Lin2022,Scammell2022}. In these twisted 2D crystals, spin fluctuations of the interaction-induced Chern bands of twisted bilayer graphene result in orbital-dependent pairing interactions~\cite{huang2022pseudospin} with a simultaneous attraction and repulsion channel, along with a significant geometric superfluid weight~\cite{hu2019geometric,xie2020topology,Tian2023}. Additionally, twisted transition metal dichalcogenides exhibit topological flat bands on bipartite honeycomb and Kagome Hubbard models~\cite{AllanTMDtwisted}, where antiferromagnetic spin fluctuations can result in orbital-dependent pairing. On the other hand, the nodal geometric BCS-PDW transition mechanism should apply to dispersive bands. In fact, PDWs have been observed in Kagome~\cite{Chen2021} and Lieb lattice superconductors~\cite{Ruan2018,Hamidian2016,Yu2019}, where the bandwidth is comparable to interactions. The possibility of band geometric instabilities in these multi-orbital superconductors is left to future studies.

\begin{acknowledgements}
The authors would like to thank Prof. E. Rossi for his comments on the manuscript. Both authors acknowledge support from UNR VPRI startup grant PG19012.
\end{acknowledgements}

\onecolumngrid
\section{S1: Calculation of the Superfluid Weight}
\label{section1}
The following supplemental sections detail the calculations for the band-projected superfluid weight in multi-orbital systems. In S1.1, the band-projected mean-field ground state energy and superfluid weight for on-site interactions are calculated. S1.2 describes the derivation of the superfluid weight formula used in S1.1. In S1.3, we extend the calculation of the superfluid weight for a general class of singlet-type interactions. There, we take a variational approach, providing an alternate derivation of the superfluid weight.

\subsection{S1.1: Calculation for the Onsite Inter-orbital Interactions}
\label{section11}
We start with the lattice Hamiltonian,
\be\label{eq:hlatticewithmu}
\ham-\mu\hat{N}_e=\sum_{ij,\alpha\beta,\sigma}h^{\sigma}_{ij,\alpha\beta}c^\dagger_{i\alpha\sigma}c_{j\beta\sigma}-\sum_{i,\alpha\beta}U_{\alpha\beta}c^\dagger_{i\alpha\uparrow}c^\dagger_{i\beta\downarrow}c_{i\beta\downarrow}c_{i\alpha\uparrow}-\mu\sum_{i\alpha\sigma}c^\dagger_{i\alpha\sigma}c_{i\alpha\sigma},
\ee
with $\hat{N}_e=\sum_{i\alpha\sigma}c^\dagger_{i\alpha\sigma}c_{i\alpha\sigma}$ the electron number operator. For $\bd{q}=0$ state, the system is spatially uniform, so one introduces order parameter matrix in orbital basis, $\hat{\Delta}_{\alpha\beta}=-U_{\alpha\beta}\la c_{i\beta\downarrow}c_{i\alpha\uparrow}\ra$,
which is independent of the site index $i$. Then the mean-field interaction can be written as
\be\label{eq:mfa}
\sum_{i,\alpha\beta}(\hat{\Delta}_{\alpha\beta} c^\dagger_{i\alpha\uparrow}c^\dagger_{i\beta\downarrow}+\hat{\Delta}^*_{\alpha\beta} c_{i\beta\downarrow}c_{i\alpha\uparrow}+\frac{|\hat{\Delta}_{\alpha\beta}|^2}{U_{\alpha\beta}}).
\ee

For the supercurrent state with $\bd{q}\neq0$, we modify $\hat{\Delta}_{\alpha\beta} \to \hat{\Delta}_{\bd{q},\alpha\beta}e^{2i\bd{q}\cdot\bd{r}_i}$ in Eq.~\ref{eq:mfa}. Then we perform a gauge transformation $c^\dagger_{i\alpha\sigma}\rightarrow c^\dagger_{i\alpha\sigma}e^{-i\bd{q}\cdot\bd{r}_i}$, which shifts the $\bd{q}$-dependent phase from the pairing term to the kinetic term \cite{peotta2015superfluidity}, yielding the $\bd{q}$-dependent mean-field Hamiltonian:
\be\label{eq:hmfq}
\begin{split}
\ham_{MF}(\bd{q})-\mu_\bd{q}\hat{N}_e=&\sum_{ij,\alpha\beta,\sigma}h^{\sigma}_{ij,\alpha\beta}e^{-i\bd{q}(\bd{r}_i-\bd{r}_j)}c^\dagger_{i\alpha\sigma}c_{j\beta\sigma}-\mu_\bd{q}\sum_{i\alpha\sigma}c^\dagger_{i\alpha\sigma}c_{i\alpha\sigma}\\
&+\sum_{i,\alpha\beta}(\hat{\Delta}_{\bd{q},\alpha\beta} c^\dagger_{i\alpha\uparrow}c^\dagger_{i\beta\downarrow}+\hat{\Delta}^*_{\bd{q},\alpha\beta} c_{i\beta\downarrow}c_{i\alpha\uparrow}+\frac{|\hat{\Delta}_{\bd{q},\alpha\beta}|^2}{U_{\alpha\beta}})\\
=&\sum_{\bd{k},\alpha\beta}\big[h^\uparrow_{\alpha\beta}(\bd{k}+\bd{q})c^\dagger_{\bd{k}\alpha\uparrow}c_{\bd{k}\beta\uparrow}+h^\downarrow_{\alpha\beta}(-\bd{k}+\bd{q})c^\dagger_{-\bd{k},\alpha\downarrow}c_{-\bd{k},\beta\downarrow}\big]-\mu_\bd{q}\sum_{\bd{k}\alpha}\big(c^\dagger_{\bd{k}\alpha\uparrow}c_{\bd{k}\alpha\uparrow}+c^\dagger_{-\bd{k},\alpha\downarrow}c_{-\bd{k},\alpha\downarrow}\big)\\
&+\sum_{\bd{k},\alpha\beta}\big(\hat{\Delta}_{\bd{q},\alpha\beta}c^\dagger_{\bd{k}\alpha\uparrow}c^\dagger_{-\bd{k},\beta\downarrow}+\hat{\Delta}_{\bd{q},\alpha\beta}^* c_{-\bd{k},\beta\downarrow}c_{\bd{k}\alpha\uparrow}+\frac{|\hat{\Delta}_{\bd{q},\alpha\beta}|^2}{U_{\alpha\beta}}\big),
\end{split}
\ee
where for the second ``=" we transform to the momentum space, and as mentioned in the main text, the modification $\mu\rightarrow\mu_\bd{q}$, $\hat{\Delta}\rightarrow\hat{\Delta}_\bd{q}$ imposes self-consistency for the particle conservation and the gap equation. After choosing the Nambu basis, Eq.~\ref{eq:hmfq} is put into the Bogoliubov-de-Genne (BdG) form,
\be\label{eq:hbdgq}
\ham_{MF}(\bd{q})-\mu_\bd{q}\hat{N}_e=\sum_\bd{k}\bd{C}^\dagger_\bd{k}H_{BdG,\bd{k}}(\bd{q})\bd{C}_\bd{k}+\sum_\bd{k}\tr\{h^\downarrow(-\bd{k}+\bd{q})-\mu_\bd{q} I_s\}+N\sum_{\alpha\beta}\frac{|\hat{\Delta}_{\bd{q},\alpha\beta}|^2}{U_{\alpha\beta}}
\ee
where
\be\label{eq:hbdgqmatrix}
H_{BdG,\bd{k}}(\bd{q})=\begin{pmatrix}
	h^\uparrow(\bd{k}+\bd{q})-\mu_\bd{q}&\hat{\Delta}_\bd{q}\\
	\hat{\Delta}^\dagger_\bd{q}&-[ h^{\downarrow T}(-\bd{k}  +\bd{q}) - \mu_\bd{q}]
\end{pmatrix},
\ee
and $\bd{C}_\bd{k}=(
c_{\bd{k}1\uparrow},..,c_{\bd{k}s\uparrow},c_{-\bd{k}1\downarrow}^\dagger,..,c_{-\bd{k}s\downarrow}^\dagger)^T$ is the $2s$-dimensional Nambu spinor, with $s$ the number of orbitals; $N$ denotes the number of unit cells, which is equal to the electron number in a filled band. Time-reversal symmetry (TRS) is imposed through $ h^{\downarrow T}(-\bd{k}) = h^\uparrow(\bd{k})$, and the general case for spin-orbit coupled systems will be discussed elsewhere.

We then project to the $m^{th}$ band, by performing transformation $G_{\bd{k}+\bd{q}}\oplus G_{\bd{k}-\bd{q}}$ to Eq.~\ref{eq:hbdgqmatrix} and collecting terms pertaining to the $m^{th}$ band only (the $l^{th}$ column of sewing matrix $G_\bd{k}$ is the Bloch function of the $l^{th}$ spin-$\uparrow$ band, $u_{l,\bd{k}}$). We obtain
\be\label{eq:hproject}
\begin{split}
\ham_{MF,m}(\bd{q})-\mu_\bd{q}\hat{N}_{e,m}=&\sum_\bd{k}\begin{pmatrix}
c^\dagger_{m,\bd{k},\uparrow}(\bd{q})&c_{m,-\bd{k},\downarrow}(\bd{q})
\end{pmatrix}\begin{pmatrix}
\xi_{m,\bd{k}+\bd{q}}&\Delta_{m,\bd{k}}(\bd{q})\\
\Delta_{m,\bd{k}}(\bd{q})^*&-\xi_{m,\bd{k}-\bd{q}}
\end{pmatrix}\begin{pmatrix}
c_{m,\bd{k},\uparrow}(\bd{q})\\c^\dagger_{m,-\bd{k},\downarrow}(\bd{q})
\end{pmatrix}\\
&+\sum_\bd{k}\xi_{m,\bd{k}-\bd{q}}+N\sum_{\alpha\beta}\frac{|\hat{\Delta}_{\bd{q},\alpha\beta}|^2}{U_{\alpha\beta}}.
\end{split}
\ee
Here $c_{m,\bd{k},\sigma}(\bd{q})$ is a new set of basis whose $\bd{q}$-dependence comes from the transformation $G_{\bd{k}\pm\bd{q}}$, and they will be the building blocks of the finite-$\bd{q}$ supercurrent state. The explicit form is
\be\label{eq:qbasis}
\begin{split}
	&c_{l\bd{k}\uparrow}(\bd{q})=\sum_\alpha u_{l,\bd{k}+\bd{q},\alpha}^*c_{\bd{k}\alpha\uparrow},\\
	&c^\dagger_{l,-\bd{k}\downarrow}(\bd{q})=\sum_\alpha u_{l,\bd{k}-\bd{q},\alpha}^*c^\dagger_{-\bd{k}\alpha\downarrow},
\end{split}
\ee
where $l$ is the band index. At $\bd{q}=0$, this is nothing but the transformation between orbital and band basis. 

Mathematically, one can verify that Eq.~\ref{eq:hproject} can be derived by acting a $\bd{q}$-dependent band projection operator
\be\label{eq:projoperator}
\hat{\mathcal{P}}_m(\bd{q})\equiv\prod_{\bd{k}}[c^\dagger_{m,\bd{k},\uparrow}(\bd{q})|0\ra \la 0|c_{m,\bd{k},\uparrow}(\bd{q})]\oplus [c^\dagger_{m,-\bd{k},\downarrow}(\bd{q})|0\ra\la0| c_{m,-\bd{k},\downarrow}(\bd{q})]
\ee
on the two sides of Eq.~\ref{eq:hmfq}: $\ham_{MF,m}(\bd{q})-\mu_\bd{q}\hat{N}_{e,m}=\hat{\mathcal{P}}_m(\bd{q})[\ham_{MF}(\bd{q})-\mu_\bd{q}\hat{N}_e]\hat{\mathcal{P}}_m(\bd{q})$. Notice we also project the electrons in the band, with $\hat{N}_{e,m}$ the number operator for electrons in the $m^{th}$ band only, since the other $s-1$ electronic bands are assumed to be either filled or empty.

The band-projected grand potential is thus defined as
\be\label{eq:grandandh}
\Omega_m(\bd{q})=-\frac{1}{\beta}\ln\tr\{e^{-\beta[\ham_{MF,m}(\bd{q})-\mu_\bd{q}\hat{N}_{e,m}]}\}.
\ee
At $T=0$, this becomes
\be\label{eq:grandpotential2}
\Omega_m(\bd{q})=\sum_\bd{k}E_{m,\bd{k},-}(\bd{q})+\sum_\bd{k}\xi_{m,\bd{k}-\bd{q}}+N\sum_{\alpha\beta}\frac{|\hat{\Delta}_{\bd{q},\alpha\beta}|^2}{U_{\alpha\beta}},
\ee
which is the main equation we use to calculate the projected superfluid weight using the grand potential formulation. $E_{m,\bd{k},\pm}(\bd{q})$ are the two quasiparticle band energy of the $m^{th}$ band:
\be
E_{m,\bd{k},\pm}(\bd{q})=\frac{1}{2}\big\{(\xi_{m,\bd{k}+\bd{q}}-\xi_{m,\bd{k}-\bd{q}})\pm\sqrt{(\xi_{m,\bd{k}+\bd{q}}+\xi_{m,\bd{k}-\bd{q}})^2+4|\Delta_{m,\bd{k}}(\bd{q})|^2}\big\}.
\ee

By definition, the band-projected superfluid weight tensor $D^m_{s,\mu\nu}$ is the total derivative of free energy with the number of electrons $N_{e,m}$ fixed. It can be shown that it is equal to the partial derivative of grand potential $\Omega_m(\bd{q},\mu_\bd{q},\hat{\Delta}_\bd{q})$ with $\mu_\bd{q}$, $\hat{\Delta}_\bd{q}$ held constant: 
\be\label{eq:dsformula}
D^m_{s,\mu\nu}\equiv\frac{1}{N}\frac{\partial^2F_m(\bd{q})}{\partial q_\mu\partial q_\nu}\bigg|_{\bd{q}=0}=\frac{1}{N}\bigg(\frac{\partial^2\Omega_m}{\partial q_\mu\partial q_\nu}\bigg)_{\mu_\bd{q}=\mu,\hat{\Delta}_\bd{q}=\hat{\Delta}}\bigg|_{\bd{q}=0}.
\ee
where $F_m(\bd{q})=\Omega_m(\bd{q},\mu_\bd{q},\hat{\Delta}_\bd{q})+\mu_\bd{q}N_{e,m}$. $\Omega_m$ is a function of $\bd{q},\mu_\bd{q},\hat{\Delta}_\bd{q}$, with the $\bd{q}$ dependence of $\mu_\bd{q},\hat{\Delta}_\bd{q}$ determined self-consistently. The derivative in Eq.~\ref{eq:dsformula} is taken with $\mu_\bd{q},\hat{\Delta}_\bd{q}$ set to $\mu,\hat{\Delta}$. The general proof for Eq.~\ref{eq:dsformula} has been given by \cite{peotta2015superfluidity,taylor2006pairing}, and we provide a derivation of this formula for our multi-orbital case in the next section.

For $D^m_{s,\mu\nu}$ to have the meaning of stability for the $\bd{q}=0$ BCS state, the first order derivative of free energy must be zero, i.e., $\bd{q}=0$ is either a saddle point or local extremum of $F_m(\bd{q})$. Defining an auxiliary quantity
\be
X_{m,\bd{k}}(\bd{q})\equiv(\xi_{m,\bd{k}+\bd{q}}+\xi_{m,\bd{k}-\bd{q}})^2+4|\Delta_{m,\bd{k}}(\bd{q})|^2,
\ee
and using Eq.~\ref{eq:grandpotential2}, one can find
\be\label{eq:1stordervanish}
\begin{split}
\frac{\partial F_m(\bd{q})}{\partial q_\mu}\bigg|_{\bd{q}=0}=\bigg(\frac{\partial\Omega_m}{\partial q_\mu}\bigg)_{\mu_\bd{q},\hat{\Delta}_\bd{q}}\bigg|_{\bd{q}=0}&=-\frac{1}{4}\sum_\bd{k}X_{m,\bd{k}}(\bd{q})^{-1/2}\tilde{\partial}_{q_\mu} X_{m,\bd{k}}(\bd{q})|_{\bd{q}=0}\\
&=-\sum_\bd{k}X_{m,\bd{k}}(\bd{q})^{-1/2}\tilde{\partial}_{q_\mu}|\Delta_{m,\bd{k}}(\bd{q})|^2|_{\bd{q}=0}.
\end{split}
\ee
Here $\tilde{\partial}_q$ means the derivative does not act on $\mu_\bd{q},\hat{\Delta}_\bd{q}$ since they have been set equal to $\mu,\hat{\Delta}$, respectively. To make Eq.~\ref{eq:1stordervanish} vanish, a sufficient condition is that when $\hat{\Delta}$ is a hermitian matrix up to an overall phase, such that $|\la u_{m,\bd{k}+\bd{q}}|\hat{\Delta}|u_{m,\bd{k}-\bd{q}}\ra|^2$ is an even function of $\bd{q}$ (this also agrees with the ``hermitian condition" in S2.2). Under this condition, Eq.~\ref{eq:dsformula} leads to the final expression
\be\label{eq:dsexpression2}
D^m_{s,\mu\nu}=\frac{1}{N}\sum_\bd{k}\bigg\{-\frac{\xi_{m,\bd{k}}}{E_{m,\bd{k}}}\partial_\mu\partial_\nu\xi_{m,\bd{k}}-\frac{\tilde{\partial}_{q_\mu}\tilde{\partial}_{q_\nu}|\Delta_{m,\bd{k}}(\bd{q})|^2|_{\bd{q}=0}}{2E_{m,\bd{k}}}\bigg\}.
\ee
The tensor $G^m_{\mu\nu}(\bd{k})=-\tilde{\partial}_{q_\mu}\tilde{\partial}_{q_\nu}|\Delta_{m,\bd{k}}(\bd{q})|^2|_{\bd{q}=0}$ has the following expression for an isolated band:
\be\label{eq:gexpression}
\begin{split}
G^m_{\mu\nu}(\bd{k})=&\bigg\{\la u_{m,\bd{k}}|\hat{\Delta}^\dagger|u_{m,\bd{k}}\ra\big[\la\partial_\mu u_{m,\bd{k}}|\hat{\Delta}|\partial_\nu u_{m,\bd{k}}\ra+\la\partial_\nu u_{m,\bd{k}}|\hat{\Delta}|\partial_\mu u_{m,\bd{k}}\ra\\
&-\la\partial_\mu\partial_\nu u_{m,\bd{k}}|\hat{\Delta}|u_{m,\bd{k}}\ra-\la u_{m,\bd{k}}|\hat{\Delta}|\partial_\mu\partial_\nu u_{m,\bd{k}}\ra\big]-\big[\la\partial_\mu u_{m,\bd{k}}|\hat{\Delta}|u_{m,\bd{k}}\ra\\
&-\la u_{m,\bd{k}}|\hat{\Delta}|\partial_\mu u_{m,\bd{k}}\ra\big]\big[\la u_{m,\bd{k}}|\hat{\Delta}^\dagger|\partial_\nu u_{m,\bd{k}}\ra-\la \partial_\nu u_{m,\bd{k}}|\hat{\Delta}^\dagger| u_{m,\bd{k}}\ra\big]\bigg\}+c.c.
\end{split}
\ee
where $\partial_\mu$ means $\partial_{k_\mu}$.

The ``conventional term" of Eq.~\ref{eq:dsexpression2}, after partial integral can be converted to
\be
-\sum_\bd{k}\frac{\xi_{m,\bd{k}}}{E_{m,\bd{k}}}\partial_\mu\partial_\nu\xi_{m,\bd{k}}=\sum_\bd{k}\frac{|\Delta_{m,\bd{k}}(0)|^2}{E_{m,\bd{k}}^3}\partial_\mu\xi_{m,\bd{k}}\partial_\nu\xi_{m,\bd{k}}-\frac{1}{4E_{m,\bd{k}}^3}\partial_\mu\xi_{m,\bd{k}}^2\partial_\nu|\Delta_{m,\bd{k}}(0)|^2,
\ee
where the first term is positive semidefinite and stabilizes the BCS state. The second term, however, is indefinite and depends on the interplay of dispersion $\xi_{m,\bd{k}}$ and geometric quantity $\partial_\nu|\Delta_{m,\bd{k}}(0)|^2$.

\subsection{S1.2: The Superfluid Weight Formula}
Here we derive the superfluid weight formula Eq.~\ref{eq:dsformula} for the band-projected case of multi-orbital systems, which is analogous to \cite{peotta2015superfluidity}. Beforehand, we show that the self-consistency equations, satisfied by $\hat{\Delta}_\bd{q},\mu_\bd{q}$, remain true under band projection. They are
\be\label{eq:cancel1}
\bigg(\frac{\partial\Omega_m}{\partial\hat{\Delta}_\bd{q}}\bigg)_{\bd{q},\mu_\bd{q}}=0
\ee
and
\be\label{eq:cancel2}
\bigg(\frac{\partial\Omega_m}{\partial\mu_\bd{q}}\bigg)_{\bd{q},\hat{\Delta}_\bd{q}}=-N_{e,m},
\ee
respectively, which hold at any $\bd{q}$ point. The first cancellation comes from the gap equation, and the second identity is by fixing electron numbers.

For a wavevector $\bd{q}\neq0$, the pairing matrix is defined as
\be\label{eq:deltaqdef}
\hat{\Delta}_{\bd{q},\alpha\beta}\equiv-U_{\alpha\beta}\la c_{i\beta\downarrow}c_{i\alpha\uparrow}\ra_\bd{q}.
\ee
Comparing with the $\bd{q}=0$ order parameter, the only change is that now the average $\la\ra_\bd{q}$ is taken with respect to the $\bd{q}\neq0$ pairing state. The form of Eq. \ref{eq:deltaqdef} is due to the gauge transformation $c^\dagger_{i\alpha\sigma}\rightarrow c^\dagger_{i\alpha\sigma}e^{-i\bd{q}\cdot\bd{r}_i}$ (see the description after Eq. \ref{eq:mfa}) and one can check the r.h.s. of Eq. \ref{eq:deltaqdef} is independent of the site index $i$. If there was not such a gauge transformation, then the quantity $-U_{\alpha\beta}\la c_{i\beta\downarrow}c_{i\alpha\uparrow}\ra_\bd{q}\propto e^{2i\bd{q}\cdot\bd{r}_i}$ and we would have to define $\hat{\Delta}_{\bd{q},\alpha\beta}\equiv-U_{\alpha\beta}\la c_{i\beta\downarrow}c_{i\alpha\uparrow}\ra_\bd{q}e^{-2i\bd{q}\cdot\bd{r}_i}$ instead.

To prove Eq.~\ref{eq:cancel1}, we first express the gap equation~\ref{eq:deltaqdef} in the band basis $c_{l\bd{k}\sigma}(\bd{q})$, and then project it to the $m^{th}$ band,
\be\label{eq:cancel1proj}
\begin{split}
\hat{\Delta}_{\bd{q},\alpha\beta}=&-\frac{U_{\alpha\beta}}{N}\sum_{\bd{k},ll'}u_{l',\bd{k}-\bd{q},\beta}^*u_{l,\bd{k}+\bd{q},\alpha}\la c_{l',-\bd{k}\downarrow}(\bd{q})c_{l\bd{k}\uparrow}(\bd{q})\ra_\bd{q}\\
\stackrel{\text{proj.}}{=}&-\frac{U_{\alpha\beta}}{N}\sum_\bd{k}u_{m,\bd{k}-\bd{q},\beta}^*u_{m,\bd{k}+\bd{q},\alpha}\la c_{m,-\bd{k}\downarrow}(\bd{q})c_{m\bd{k}\uparrow}(\bd{q})\ra_\bd{q}.
\end{split}
\ee
Since we assume that the $m^{th}$ band is isolated from the other bands by a large energy gap, the field components of $m^{th}$ band dominate in the supercurrent state. Therefore, the fields from other bands can be ignored, justifying our projection.

Next, using the grand potential expression Eq.~\ref{eq:grandandh}, we find at finite temperature
\be\label{eq:cancel1proof}
\begin{split}
\bigg(\frac{\partial\Omega_m}{\partial\hat{\Delta}_{\bd{q},\alpha\beta}}\bigg)_{\bd{q},\mu_\bd{q}}&=\frac{1}{\tr\{e^{-\beta[\ham_{MF,m}(\bd{q})-\mu_\bd{q}\hat{N}_{e,m}]}\}}\tr\bigg\{e^{-\beta[\ham_{MF,m}(\bd{q})-\mu_\bd{q}\hat{N}_{e,m}]}\bigg(\frac{\partial[\ham_{MF,m}(\bd{q})-\mu_\bd{q}\hat{N}_{e,m}]}{\partial\hat{\Delta}_{\bd{q},\alpha\beta}}\bigg)_{\bd{q},\mu_\bd{q}}\bigg\}\\
&=\sum_\bd{k}\bigg(\frac{\partial\Delta_{m,\bd{k}}(\bd{q})}{\partial\hat{\Delta}_{\bd{q},\alpha\beta}}\bigg)_{\bd{q}}\la c^\dagger_{m\bd{k}\uparrow}(\bd{q})c^\dagger_{m,-\bd{k}\downarrow}(\bd{q})\ra_\bd{q}+\frac{N}{U_{\alpha\beta}}\hat{\Delta}_{\bd{q},\alpha\beta}^*\\
&=\sum_\bd{k}u_{m,\bd{k}+\bd{q},\alpha}^*u_{m,\bd{k}-\bd{q},\beta}\la c^\dagger_{m\bd{k}\uparrow}(\bd{q})c^\dagger_{m,-\bd{k}\downarrow}(\bd{q})\ra_\bd{q}+\frac{N}{U_{\alpha\beta}}\hat{\Delta}_{\bd{q},\alpha\beta}^*.
\end{split}
\ee
This is exactly the complex conjugate of Eq.~\ref{eq:cancel1proj}, so Eq.~\ref{eq:cancel1} is proved. Here, $\hat{\Delta}_{\bd{q},\alpha\beta}$ and $\hat{\Delta}_{\bd{q},\alpha\beta}^*$ are treated as independent variables.

To prove Eq.~\ref{eq:cancel2}, we write down the total number of electrons
\be
N_e=\sum_{i,\alpha,\sigma}\la c^\dagger_{i\alpha\sigma}c_{i\alpha\sigma}\ra_\bd{q}
=\sum_{ll',\bd{k},\alpha}u_{l,\bd{k}+\bd{q},\alpha}^*u_{l',\bd{k}+\bd{q},\alpha}\la c^\dagger_{l\bd{k}\uparrow}(\bd{q})c_{l'\bd{k}\uparrow}(\bd{q})\ra_\bd{q}+u_{l,\bd{k}-\bd{q},\alpha}u_{l',\bd{k}-\bd{q},\alpha}^*\la c^\dagger_{l,-\bd{k}\downarrow}(\bd{q})c_{l',-\bd{k}\downarrow}(\bd{q})\ra_\bd{q}.
\ee
By band projection,
\be\label{eq:cancel2proj}
\begin{split}
N_{e,m}&=\sum_{\bd{k},\alpha}|u_{m,\bd{k}+\bd{q},\alpha}|^2\la c^\dagger_{m\bd{k}\uparrow}(\bd{q})c_{m\bd{k}\uparrow}(\bd{q})\ra_\bd{q}+|u_{m,\bd{k}-\bd{q},\alpha}|^2\la c^\dagger_{m,-\bd{k}\downarrow}(\bd{q})c_{m,-\bd{k}\downarrow}(\bd{q})\ra_\bd{q}\\
&=\sum_{\bd{k}}\la c^\dagger_{m\bd{k}\uparrow}(\bd{q})c_{m\bd{k}\uparrow}(\bd{q})\ra_\bd{q}+\la c^\dagger_{m,-\bd{k}\downarrow}(\bd{q})c_{m,-\bd{k}\downarrow}(\bd{q})\ra_\bd{q},
\end{split}
\ee
where in the second line we used normalization $\sum_\alpha|u_{m,\bd{k},\alpha}|^2=1$.

On the other hand, from Eq.~\ref{eq:grandandh},
\be\label{eq:cancel2proof}
\begin{split}
\bigg(\frac{\partial\Omega_m}{\partial\mu_\bd{q}}\bigg)_{\bd{q},\hat{\Delta}_\bd{q}}&=\frac{1}{\tr\{e^{-\beta[\ham_{MF,m}(\bd{q})-\mu_\bd{q}\hat{N}_{e,m}]}\}}\tr\bigg\{e^{-\beta[\ham_{MF,m}(\bd{q})-\mu_\bd{q}\hat{N}_{e,m}]}\bigg(\frac{\partial[\ham_{MF,m}(\bd{q})-\mu_\bd{q}\hat{N}_{e,m}]}{\partial\mu_\bd{q}}\bigg)_{\bd{q},\hat{\Delta}_\bd{q}}\bigg\}\\
&=\sum_\bd{k}[-\la c^\dagger_{m\bd{k}\uparrow}(\bd{q})c_{m\bd{k}\uparrow}(\bd{q})\ra_\bd{q}+\la c_{m,-\bd{k}\downarrow}(\bd{q})c^\dagger_{m,-\bd{k}\downarrow}(\bd{q})\ra_\bd{q}]-N,
\end{split}
\ee
which is exactly the minus Eq.~\ref{eq:cancel2proj}, so Eq.~\ref{eq:cancel2} is proved.

With Eq.~\ref{eq:cancel1} and Eq.~\ref{eq:cancel2} established, using $F_m(\bd{q})=\Omega_m(\bd{q},\mu_\bd{q},\hat{\Delta}_{\bd{q},\alpha\beta})+\mu_\bd{q}N_{e,m}$, for the first-order derivative we find
\be\label{eq:1storder}
\frac{\partial F_m}{\partial q_\mu}=\bigg(\frac{\partial \Omega_m}{\partial q_\mu}\bigg)_{\mu_\bd{q},\hat{\Delta}_{\bd{q},\alpha\beta}}+\bigg(\frac{\partial \Omega_m}{\partial\mu_\bd{q}}\bigg)_{\bd{q},\hat{\Delta}_{\bd{q},\alpha\beta}}\frac{\partial\mu_\bd{q}}{\partial q_\mu}+\bigg(\frac{\partial \Omega_m}{\partial\hat{\Delta}_{\bd{q},\alpha\beta}}\bigg)_{\bd{q},\mu_{\bd{q}}}\frac{\partial\hat{\Delta}_{\bd{q},\alpha\beta}}{\partial q_\mu}+\frac{\partial\mu_\bd{q}}{\partial q_\mu}N_{e,m}=\bigg(\frac{\partial \Omega_m}{\partial q_\mu}\bigg)_{\mu_\bd{q},\hat{\Delta}_{\bd{q},\alpha\beta}},
\ee
and for the second-order derivative,
\be\label{eq:lastvanish}
\frac{\partial^2 F_m}{\partial q_\mu\partial q_\nu}\bigg|_{\bd{q}=0}=\bigg(\frac{\partial^2 \Omega_m}{\partial q_\mu\partial q_\nu}\bigg)_{\mu_\bd{q},\hat{\Delta}_{\bd{q}}}\bigg|_{\bd{q}=0}+\bigg(\frac{\partial^2 \Omega_m}{\partial q_\mu\partial\mu_\bd{q}}\bigg)_{\hat{\Delta}_{\bd{q}}}\bigg|_{\bd{q}=0}\frac{\partial\mu_\bd{q}}{\partial q_\mu}\bigg|_{\bd{q}=0}+\sum_{\alpha\beta}\bigg(\frac{\partial^2 \Omega_m}{\partial q_\mu\partial\hat{\Delta}_{\bd{q},\alpha\beta}}\bigg)_{\mu_\bd{q}}\bigg|_{\bd{q}=0}\frac{\partial\hat{\Delta}_{\bd{q},\alpha\beta}}{\partial q_\mu}\bigg|_{\bd{q}=0}.
\ee
In a later section, S2.2, we will show that for the multi-orbital case, $\mu_\bd{q}$ and each matrix element $\hat{\Delta}_{\bd{q},\alpha\beta}$ (under a ``hermitian condition") are both even functions of $\bd{q}$, so the last two terms vanish. Therefore we arrive at Eq.~\ref{eq:dsformula}.
\be
D^m_{s,\mu\nu}\equiv\frac{1}{N}\frac{\partial^2F_m(\bd{q})}{\partial q_\mu\partial q_\nu}\bigg|_{\bd{q}=0}=\frac{1}{N}\bigg(\frac{\partial^2\Omega_m}{\partial q_\mu\partial q_\nu}\bigg)_{\mu_\bd{q}=\mu,\hat{\Delta}_\bd{q}=\hat{\Delta}}\bigg|_{\bd{q}=0}.
\ee

\subsection{S1.3: Calculation for the General Singlet-type Interactions (Variational Method)}
\label{section13}
Here we establish the band-projection formalism for general singlet-type interactions. The interaction can be between Cooper pairs of any orbitals and across any distance in the lattice (which is beyond density-density interactions and includes exchange interactions), so it is denoted by matrix elements $V^{(\alpha\beta,\gamma\delta)}_{\bd{k}\bd{k}'}$. In principle, the problem can be solved using the same BdG approach as in the main text; however, we find it also insightful to solve with the variational method. We will derive a self-consistency equation for the band-projected gap function $\Delta_{m,\bd{k}}(\bd{q})$, which can be simplified by an expansion of the interaction into irreducible representations of the lattice point groups. This eventually leads to an analytical expression for the superfluid weight.

For the $\bd{q}\neq0$ supercurrent state, one can start from a lattice hamiltonian similar to Eq.~1 of the main text, but look at a $\bd{q}$-slice of the reduced interaction $\sum_{\bd{k}\bd{k}',\alpha\beta\gamma\delta}V^{(\alpha\beta,\gamma\delta)}_{\bd{k}\bd{k}'}c^\dagger_{\bd{k}+\bd{q},\alpha\uparrow}c^\dagger_{-\bd{k}+\bd{q},\beta\downarrow}c_{-\bd{k}+\bd{q},\delta\downarrow}c_{\bd{k}+\bd{q},\gamma\uparrow}$. Then one can either right away proceed with the variational method, or do a gauge transformation $c^\dagger_{i\alpha\sigma}\rightarrow c^\dagger_{i\alpha\sigma}e^{-i\bd{q}\cdot\bd{r}_i}$ to shift the $\bd{q}$-dependence to the kinetic part. Here we take the second way to get consistent with the main text. The $\bd{q}$-dependent reduced Hamiltonian is
\be\label{eq:hq}
\ham_\text{red}(\bd{q})-\mu_\bd{q}\hat{N}_e=\sum_{\bd{k},\alpha\beta,\sigma}c^\dagger_{\bd{k}\alpha\sigma}h^\sigma_{\alpha\beta}(\bd{k}+\bd{q})c_{\bd{k}\beta\sigma}+\sum_{\bd{k}\bd{k}',\alpha\beta\gamma\delta}V^{(\alpha\beta,\gamma\delta)}_{\bd{k}\bd{k}'}c^\dagger_{\bd{k}\alpha\uparrow}c^\dagger_{-\bd{k}\beta\downarrow}c_{-\bd{k}'\delta\downarrow}c_{\bd{k}'\gamma\uparrow}-\mu_\bd{q}\sum_{\bd{k}\alpha\sigma}c^\dagger_{\bd{k}\alpha\sigma}c_{\bd{k}\alpha\sigma}.
\ee
Recall that $h^\sigma_{\alpha\beta}(\bd{k})$ for $\sigma=\uparrow,\downarrow$ can be diagonalized by matrices $G_\bd{k}$ and $G_{-\bd{k}}^*$, respectively. This implies that to project to the $m^{th}$ band, we need to go to the $c_{l\bd{k}\sigma}(\bd{q})$ basis in Eq.~\ref{eq:qbasis}. Rewrite the hamiltonian into this basis, and then project to the $m^{th}$ band, obtaining
\be\label{eq:hmproject}
\begin{split}
	\ham_{\text{red},m}(\bd{q})-\mu_\bd{q}\hat{N}_{e,m}=&\sum_\bd{k}c^\dagger_{m\bd{k}\uparrow}(\bd{q})\xi_{m,\bd{k}+\bd{q}}c_{m\bd{k}\uparrow}(\bd{q})+c^\dagger_{m,-\bd{k}\downarrow}(\bd{q})\xi_{m,\bd{k}-\bd{q}}c_{m,-\bd{k}\downarrow}(\bd{q})\\
	&+\sum_{\bd{k}\bd{k}'}V^{(m)}_{\bd{k}\bd{k}'}(\bd{q})c^\dagger_{m\bd{k}\uparrow}(\bd{q})c^\dagger_{m,-\bd{k}\downarrow}(\bd{q})c_{m,-\bd{k}'\downarrow}(\bd{q})c_{m\bd{k}'\uparrow}(\bd{q}),
\end{split}
\ee
where we defined the \tit{$\bd{q}$-dependent band-projected interaction}
\be\label{eq:vmqdef}
V^{(m)}_{\bd{k}\bd{k}'}(\bd{q})\equiv\sum_{\alpha\beta\gamma\delta}V^{(\alpha\beta,\gamma\delta)}_{\bd{k}\bd{k}'}u_{m,\bd{k}+\bd{q},\alpha}^*u_{m,\bd{k}-\bd{q},\beta}u_{m,\bd{k}'-\bd{q},\delta}^*u_{m,\bd{k}'+\bd{q},\gamma}.
\ee
One can also verify that the band-projected hamiltonian Eq.~\ref{eq:hmproject} is related to Eq.~\ref{eq:hq} through $\ham_{\text{red},m}(\bd{q})-\mu_\bd{q}\hat{N}_{e,m}=\hat{\mathcal{P}}_m(\bd{q})[\ham_\text{red}(\bd{q})-\mu_\bd{q}\hat{N}_e]\hat{\mathcal{P}}_m(\bd{q})$, with $\hat{\mathcal{P}}_m(\bd{q})$ given by Eq.~\ref{eq:projoperator}.

Hamiltonian Eq.~\ref{eq:hmproject} indicates that the mean-field pairing ground state has the ansatz form
\be\label{eq:groundq}
|\Psi_G(\bd{q})\ra=\prod_\bd{k}[w_{m\bd{k}}(\bd{q})e^{i\varphi_{m\bd{k}}(\bd{q})}+v_{m\bd{k}}(\bd{q})c^\dagger_{m\bd{k}\uparrow}(\bd{q})c^\dagger_{m,-\bd{k}\downarrow}(\bd{q})]|0\ra,
\ee
where $w_{m\bd{k}}(\bd{q}),v_{m\bd{k}}(\bd{q})$ and $\varphi_{m\bd{k}}(\bd{q})$ are real parameters to be determined from the minimization of the energy functional $E_G(\bd{q})=\la\Psi_G(\bd{q})|\ham_{\text{red},m}(\bd{q})-\mu_\bd{q}\hat{N}_{e,m}|\Psi_G(\bd{q})\ra$.

The ground state $\Psi_G(\bd{q})$ needs to be normalized, so one can introduce parameter $\theta_{m\bd{k}}(\bd{q})$ by $w_{m\bd{k}}(\bd{q})=\cos\theta_{m\bd{k}}(\bd{q})$ and $v_{m\bd{k}}(\bd{q})=\sin\theta_{m\bd{k}}(\bd{q})$. In solving the variational equations with respect to $\theta_{m\bd{k}}(\bd{q})$ and $\varphi_{m\bd{k}}(\bd{q})$, it is useful to define the band-projected gap function
\be\label{eq:gapm2}
\Delta_{m,\bd{k}}(\bd{q})=\frac{1}{2}\sum_{\bd{k}'}V^{(m)}_{\bd{k}\bd{k}'}(\bd{q})\sin2\theta_{m\bd{k}'}(\bd{q})e^{-i\varphi_{m\bd{k}'}(\bd{q})},
\ee
so the variational equations lead to a self-consistency equation
\be\label{eq:self}
\Delta_{m,\bd{k}}(\bd{q})=-\sum_{\bd{k}'}V^{(m)}_{\bd{k}\bd{k}'}(\bd{q})\frac{\Delta_{m,\bd{k}'}(\bd{q})}{\sqrt{(\xi_{m,\bd{k}'+\bd{q}}+\xi_{m,\bd{k}'-\bd{q}})^2+4|\Delta_{m,\bd{k}'}(\bd{q})|^2}}.
\ee
It turns out that this $\Delta_{m,\bd{k}}(\bd{q})$ is identical to the definition given in the main text, thus Eq.~\ref{eq:self} is the self-consistency equation that the band-projected gap function $\Delta_{m,\bd{k}}(\bd{q})$ must satisfy.

In terms of $\Delta_{m,\bd{k}}(\bd{q})$, the minimized ground state energy is found to be
\be\label{eq:egq}
\begin{split}
	E_G(\bd{q})=&\sum_\bd{k}\frac{1}{2}\bigg\{(\xi_{m,\bd{k}+\bd{q}}-\xi_{m,\bd{k}-\bd{q}})-\sqrt{(\xi_{m,\bd{k}+\bd{q}}+\xi_{m,\bd{k}-\bd{q}})^2+4|\Delta_{m,\bd{k}}(\bd{q})|^2}\bigg\}\\
	&+\sum_\bd{k}\xi_{m,\bd{k}-\bd{q}}+\sum_\bd{k}\frac{|\Delta_{m,\bd{k}}(\bd{q})|^2}{\sqrt{(\xi_{m,\bd{k}+\bd{q}}+\xi_{m,\bd{k}-\bd{q}})^2+4|\Delta_{m,\bd{k}}(\bd{q})|^2}}.
\end{split}
\ee
Expression Eq.~\ref{eq:egq} is identical to the band-projected grand potential Eq.~\ref{eq:grandpotential2}, but with the last term modified for the general type of interactions. We show below that the superfluid weight at $T=0$ can be calculated using formula $D^m_{s,\mu\nu}=(1/N)\tilde{\partial}_\mu\tilde{\partial}_\nu E_G(\bd{q})|_{\bd{q}=0}$. We will unravel the role of order parameters here by an irreducible expansion of the interaction $V^{(\alpha\beta,\gamma\delta)}_{\bd{k}\bd{k}'}$.

\subsection{Irreducible Expansion of the Interaction}
The structure of $V^{(m)}_{\bd{k}\bd{k}'}(\bd{q})$ suggests that the solution to Eq.~\ref{eq:self} can be cast into the form
\be\label{eq:gapm2}
\Delta_{m,\bd{k}}(\bd{q})=\la u_{m,\bd{k}+\bd{q}}|\hat{\Delta}(\bd{k})_\bd{q}|u_{m,\bd{k}-\bd{q}}\ra,
\ee
with $\hat{\Delta}(\bd{k})_\bd{q}$ a generalization of $\hat{\Delta}_{\bd{q}}$ for non-$s$-wave pairing. Then one would find that Eq.~\ref{eq:self} reduces to a self-consistency equation about $\hat{\Delta}(\bd{k})_\bd{q}$:
\be\label{eq:deltaqself1}
\hat{\Delta}(\bd{k})_{\bd{q},\alpha\beta}=\sum_{\bd{k}',\gamma\delta}V^{(\alpha\beta,\gamma\delta)}_{\bd{k}\bd{k}'}\la c_{-\bd{k}'\delta\downarrow}c_{\bd{k}'\gamma\uparrow}\ra_\bd{q},
\ee
where $\la\ra_\bd{q}$ denotes the average under $\Psi_G(\bd{q})$ and
\be
\la c_{-\bd{k}'\delta\downarrow}c_{\bd{k}'\gamma\uparrow}\ra_\bd{q}=-u_{m,\bd{k}'-\bd{q},\delta}^*u_{m,\bd{k}'+\bd{q},\gamma}\frac{\Delta_{m,\bd{k}'}(\bd{q})}{\sqrt{(\xi_{m,\bd{k}'+\bd{q}}+\xi_{m,\bd{k}'-\bd{q}})^2+4|\Delta_{m,\bd{k}'}(\bd{q})|^2}}.
\ee
In the particular case of onsite Hubbard interactions, $V^{(\alpha\beta,\gamma\delta)}_{\bd{k}\bd{k}'}=-(1/N)U_{\alpha\beta}\delta_{\alpha\gamma}\delta_{\beta\delta}$ has no $\bd{k},\bd{k}'$-dependence, then one can check that these are consistent with Eq.~\ref{eq:deltaqdef}.

To solve Eq.~\ref{eq:deltaqself1} for the general case, we expand $V^{(\alpha\beta,\gamma\delta)}_{\bd{k}\bd{k}'}$ into irreducible representations of the lattice point group:
\be\label{eq:expansion}
V^{(\alpha\beta,\gamma\delta)}_{\bd{k}\bd{k}'}=-\sum_\Gamma v^{(\alpha\beta,\gamma\delta)}_\Gamma\chi^{(\alpha\beta)*}_\Gamma(\bd{k})\chi^{(\gamma\delta)}_\Gamma(\bd{k}'),
\ee
where $\chi^{(\alpha\beta)}_\Gamma(\bd{k})$ is the form factor of irreducible representation $\Gamma$, and $v^{(\alpha\beta,\gamma\delta)}_\Gamma$ are real coefficients. Combining Eq.~\ref{eq:deltaqself1} to \ref{eq:expansion}, suppose we have successfully solved the leading channel $\Gamma_0$ that minimizes the ground state energy, then the last term of $E_G(\bd{q})$ in Eq.~\ref{eq:egq} has a simple expression
\be\label{eq:lastterm}
\begin{split}
\sum_\bd{k}\frac{|\Delta_{m,\bd{k}}(\bd{q})|^2}{\sqrt{(\xi_{m,\bd{k}-\bd{q}}+\xi_{m,\bd{k}+\bd{q}})^2+4|\Delta_{m,\bd{k}}(\bd{q})|^2}}=&\sum_{\bd{k}\bd{k}'}\sum_{\alpha\beta\gamma\delta}v^{(\alpha\beta,\gamma\delta)}_{\Gamma_0}\chi^{(\alpha\beta)}_{\Gamma_0}(\bd{k})\chi^{(\gamma\delta)*}_{\Gamma_0}(\bd{k}')\la c_{-\bd{k}\beta\downarrow}c_{\bd{k}\alpha\uparrow}\ra_\bd{q}\la c^\dagger_{\bd{k}'\gamma\uparrow}c^\dagger_{-\bd{k}'\delta\downarrow}\ra_\bd{q}\\
=&\sum_{\alpha\beta\gamma\delta}v^{(\alpha\beta,\gamma\delta)}_{\Gamma_0}\tilde{\Delta}^{(\alpha\beta)}_{\Gamma_0}(\bd{q})\tilde{\Delta}^{(\gamma\delta)}_{\Gamma_0}(\bd{q})^*.
\end{split}
\ee
where we defined an auxiliary order parameter matrix
\be\label{eq:neworder}
\tilde{\Delta}^{(\alpha\beta)}_{\Gamma_0}(\bd{q})\equiv-\sum_{\bd{k}}\chi^{(\alpha\beta)}_{\Gamma_0}(\bd{k})\la c_{-\bd{k},\beta\downarrow}c_{\bd{k}\alpha\uparrow}\ra_\bd{q},
\ee
The l.h.s. of Eq.~\ref{eq:lastterm} is positive, which puts constraints on the leading channel $\Gamma_0$, e.g., when there is only one orbital, $v_{\Gamma_0}$ has to be positive, so the interaction must be attractive. In other words, the coefficients $v^{(\alpha\beta,\gamma\delta)}_{\Gamma_0}$ determine whether the ground state is superconducting or not. Eq.~\ref{eq:lastterm} also indicates that $\tilde{\Delta}_{\Gamma_0}(\bd{q})$ will play the same role as $\hat{\Delta}_{\bd{q}}$ playing in our earlier calculations.

Indeed, if we write $E_G$ as a function of $\bd{q}$, $\mu_\bd{q}$, $\tilde{\Delta}^{(\alpha\beta)}_{\Gamma_0}(\bd{q})$ and $\tilde{\Delta}^{(\gamma\delta)}_{\Gamma_0}(\bd{q})^*$, using definition $X_{m,\bd{k}}(\bd{q})\equiv(\xi_{m,\bd{k}-\bd{q}}+\xi_{m,\bd{k}+\bd{q}})^2+4|\Delta_{m,\bd{k}}(\bd{q})|^2$, one can verify that
\be
\begin{split}
\bigg(\frac{\partial E_G}{\partial \tilde{\Delta}^{(\alpha\beta)}_{\Gamma_0}(\bd{q})}\bigg)_{\bd{q},\mu_\bd{q},\tilde{\Delta}^{(\gamma\delta)}_{\Gamma_0}(\bd{q})^*}=&-\frac{1}{2}\sum_\bd{k}\bigg(\frac{\partial X_{m,\bd{k}}(\bd{q})^{1/2}}{\partial \tilde{\Delta}^{(\alpha\beta)}_{\Gamma_0}(\bd{q})}\bigg)_{\bd{q},\tilde{\Delta}^{(\gamma\delta)}_{\Gamma_0}(\bd{q})^*}+\sum_{\gamma\delta}v^{(\alpha\beta,\gamma\delta)}_{\Gamma_0}\tilde{\Delta}^{(\gamma\delta)}_{\Gamma_0}(\bd{q})^*\\
=&\sum_{\bd{k},\gamma\delta}v^{(\alpha\beta,\gamma\delta)}_{\Gamma_0}\chi^{(\gamma\delta)}_{\Gamma_0}(\bd{k})^*\la c^\dagger_{\bd{k}\gamma\uparrow}c^\dagger_{-\bd{k}\delta\downarrow}\ra_\bd{q}+\sum_{\gamma\delta}v^{(\alpha\beta,\gamma\delta)}_{\Gamma_0}\tilde{\Delta}^{(\gamma\delta)}_{\Gamma_0}(\bd{q})^*=0
\end{split}
\ee
for any $\bd{q}$, where the cancellation is by definition Eq.~\ref{eq:neworder}. The choice of auxiliary order parameters simplifies the calculation of superfluid weight, and we are led to
\be\label{eq:ds2}
D^m_{s,\mu\nu}=\frac{1}{N}\bigg(\frac{\partial^2 E_G}{\partial q_\mu\partial q_\nu}\bigg)_{\mu_\bd{q},\tilde{\Delta}^{(\alpha\beta)}_{\Gamma_0}(\bd{q}),\tilde{\Delta}^{(\gamma\delta)}_{\Gamma_0}(\bd{q})^*}\bigg|_{\bd{q}=0},
\ee
where by holding $\tilde{\Delta}^{(\gamma\delta)}_{\Gamma_0}(\bd{q})$, we also hold $\hat{\Delta}(\bd{k})_\bd{q}$, since
\be
\hat{\Delta}(\bd{k})_{\bd{q},\alpha\beta}=\sum_{\gamma\delta} v_{\Gamma_0}^{(\alpha\beta,\gamma\delta)}\chi^{(\alpha\beta)*}_{\Gamma_0}(\bd{k})\tilde{\Delta}^{(\gamma\delta)}_{\Gamma_0}(\bd{q}).
\ee

As a result, we obtain the expression of superfluid weight
\be\label{eq:ds3}
D^m_{s,\mu\nu}=\frac{1}{N}\sum_\bd{k}\bigg\{-\frac{\xi_{m,\bd{k}}}{E_{m,\bd{k}}}\partial_\mu\partial_\nu\xi_{{m},\bd{k}}+\frac{G^m_{\mu\nu}(\bd{k})}{2E_{m,\bd{k}}}\bigg\},
\ee
which is an extension of the formula in the main text. Here $\hat{\Delta}_\bd{q}$ and $\hat{\Delta}$ have been replaced with $\hat{\Delta}(\bd{k})_\bd{q}$ and  $\hat{\Delta}(\bd{k})$ everywhere, which are solutions for the leading channel $\Gamma_0$.


\section{S2: Symmetries of the Self-consistency Equations}
In this section, we analyze the symmetry properties of self-consistency equations. In S4.1, we give the self-consistency equations explicitly for a projected band, which can be directly applied to numerical calculations. In S4.2, we show these equations have symmetry between $\bd{q}$ and $-\bd{q}$ state. In S4.3, we show the periodicity of time-reversal invariant momentum (TRIM) in $\bd{q}$ space, which the TRS imposes.
\subsection{S2.1: The Self-consistency Equations}
For the general singlet-type interactions, the grand potential at $T=0$ is (see Eq.~\ref{eq:egq})
\be\label{eq:grandpotentialnew}
\begin{split}
\Omega_m(\bd{q})=&\sum_\bd{k}E_{m,\bd{k},-}(\bd{q})+\sum_\bd{k}\xi_{m,\bd{k}-\bd{q}}+\sum_\bd{k}\frac{|\Delta_{m,\bd{k}}(\bd{q})|^2}{\sqrt{(\xi_{m,\bd{k}+\bd{q}}+\xi_{m,\bd{k}-\bd{q}})^2+4|\Delta_{m,\bd{k}}(\bd{q})|^2}},\\
=&\sum_\bd{k}E_{m,\bd{k},-}(\bd{q})+\sum_\bd{k}\xi_{m,\bd{k}-\bd{q}}+\sum_{\alpha\beta\gamma\delta}v^{(\alpha\beta,\gamma\delta)}_{\Gamma_0}\tilde{\Delta}^{(\alpha\beta)}_{\Gamma_0}(\bd{q})\tilde{\Delta}^{(\gamma\delta)}_{\Gamma_0}(\bd{q})^*.
\end{split}
\ee
We write $\Omega_m$ as $\Omega_m(\bd{q},\mu_\bd{q},\tilde{\Delta}^{(\alpha\beta)}_{\Gamma_0}(\bd{q}))$, so the self-consistency equations Eq.~\ref{eq:cancel1}, \ref{eq:cancel2} are modified to
\be\label{eq:cancelnew}
\bigg(\frac{\partial\Omega_m}{\partial\tilde{\Delta}^{(\alpha\beta)}_{\Gamma_0}(\bd{q})}\bigg)_{\bd{q},\mu_\bd{q}}=0,\,\,\,\,\,\,\,\,\,\bigg(\frac{\partial\Omega_m}{\partial\mu_\bd{q}}\bigg)_{\bd{q},\tilde{\Delta}^{(\alpha\beta)}_{\Gamma_0}(\bd{q})}=-N_{e,m}.
\ee
Plugging Eq.~\ref{eq:grandpotentialnew} in, they are
\be\label{eq:deltaperiod1}
\hat{\Delta}(\bd{k})_{\bd{q},\alpha\beta}=-\sum_{\bd{k}',\gamma\delta}V^{(\alpha\beta,\gamma\delta)}_{\bd{k}\bd{k}'}u_{m,\bd{k}'-\bd{q},\delta}^*u_{m,\bd{k}'+\bd{q},\gamma}\frac{\Delta_{m,\bd{k}'}(\bd{q})}{\sqrt{(\varepsilon_{m,\bd{k}'+\bd{q}}+\varepsilon_{m,\bd{k}'-\bd{q}}-2\mu_\bd{q})^2+4|\Delta_{m,\bd{k}'}(\bd{q})|^2}},
\ee
\be\label{eq:muperiod1}
\sum_{\bd{k}'}\frac{\varepsilon_{m,\bd{k}'+\bd{q}}+\varepsilon_{m,\bd{k}'-\bd{q}}-2\mu_\bd{q}}{\sqrt{(\varepsilon_{m,\bd{k}'+\bd{q}}+\varepsilon_{m,\bd{k}'-\bd{q}}-2\mu_\bd{q})^2+4|\Delta_{m,\bd{k}'}(\bd{q})|^2}}-N=-N_{e,m}.
\ee
Eq.~\ref{eq:deltaperiod1} and \ref{eq:muperiod1} are the main self-consistency equations that we need to solve for a projected band. One can check that Eq.~\ref{eq:deltaperiod1} is the same as Eq.~\ref{eq:self}, with $\Delta_{m,\bd{k}'}(\bd{q})\equiv\la u_{m,\bd{k}'+\bd{q}}|\hat{\Delta}(\bd{k}')_{\bd{q}}|u_{m,\bd{k}'-\bd{q}}\ra$.

\subsection{S2.2: Symmetry between the $\bd{q}$ and $-\bd{q}$ State}
Now we prove that in the presence of TRS, the self-consistency solutions have symmetry $\mu_{-\bd{q}} = \mu_{\bd{q}}$ and $\hat{\Delta}(\bd{k})_{-\bd{q}}= \hat{\Delta}(\bd{k})_{\bd{q}}^{\dagger}$, i.e., if $\mu_\bd{q}$ and $\hat{\Delta}(\bd{k})_\bd{q}$ are solutions for the $\bd{q}$ state, then $\mu_\bd{q}$ and $\hat{\Delta}(\bd{k})_\bd{q}^{\dagger}$ is a solution for the $-\bd{q}$ state. A consequence of this is that both the band-projected grand potential $\Omega_m(\bd{q})$ and free energy $F_m(\bd{q})$ are even functions of $\bd{q}$.

To see this, for Eq.~\ref{eq:deltaperiod1}, we change $\bd{q}\rightarrow-\bd{q}$, $\alpha\leftrightarrow\beta$ and take the complex conjugate; for Eq.~\ref{eq:muperiod1}, we change $\bd{q}\rightarrow-\bd{q}$, then
\be
\hat{\Delta}(\bd{k})_{-\bd{q},\beta\alpha}^*=-\sum_{\bd{k}',\gamma\delta}V^{(\beta\alpha,\gamma\delta)}_{\bd{k}\bd{k}'}u_{m,\bd{k}'+\bd{q},\delta}u_{m,\bd{k}'-\bd{q},\gamma}^*\frac{\Delta_{m,\bd{k}'}(-\bd{q})^*}{\sqrt{(\varepsilon_{m,\bd{k}'+\bd{q}}+\varepsilon_{m,\bd{k}'-\bd{q}}-2\mu_{-\bd{q}})^2+4|\Delta_{m,\bd{k}'}(-\bd{q})|^2}},
\ee
\be
\sum_{\bd{k}'}\frac{\varepsilon_{m,\bd{k}'+\bd{q}}+\varepsilon_{m,\bd{k}'-\bd{q}}-2\mu_{-\bd{q}}}{\sqrt{(\varepsilon_{m,\bd{k}'+\bd{q}}+\varepsilon_{m,\bd{k}'-\bd{q}}-2\mu_{-\bd{q}})^2+4|\Delta_{m,\bd{k}'}(-\bd{q})|^2}}-N=-N_{e,m}.
\ee
Notice that $V^{(\beta\alpha,\gamma\delta)}_{\bd{k}\bd{k}'}=V^{(\alpha\beta,\delta\gamma)}_{\bd{k}\bd{k}'}$ by TRS, and $\Delta_{m,\bd{k}'}(-\bd{q})^*=\la u_{m,\bd{k}'+\bd{q}}|\hat{\Delta}(\bd{k}')_{-\bd{q}}^\dagger|u_{m,\bd{k}'-\bd{q}}\ra$. The two equations above say that $\mu_{-\bd{q}}$ and $\hat{\Delta}(\bd{k})_{-\bd{q}}^\dagger$ satisfy the same set of self-consistency equations that $\mu_\bd{q}$ and $\hat{\Delta}(\bd{k})_\bd{q}$ satisfy, therefore our assertion is proved. Using Eq.~\ref{eq:grandpotentialnew} and $F_m(\bd{q})=\Omega_m(\bd{q})+\mu_\bd{q}N_e$, one can further show $\Omega_m(-\bd{q})=\Omega_m(\bd{q})$ and $F_m(-\bd{q})=F_m(\bd{q})$.

Now that we know $\hat{\Delta}_{-\bd{q}}=\hat{\Delta}_\bd{q}^\dagger$, if we further assume that it satisfies a ``hermitian condition" that \tit{for any small $\bd{q}$, by a gauge choice the solution $\hat{\Delta}_{\bd{q}}$ can be made (approximately) hermitian}, then we simultaneously have 
$\hat{\Delta}_{-\bd{q},\alpha\beta}=\hat{\Delta}_{\bd{q},\beta\alpha}^*$ and $\hat{\Delta}_{\bd{q},\alpha\beta}=\hat{\Delta}_{\bd{q},\beta\alpha}^*$, which means each matrix element $\hat{\Delta}_{\bd{q},\alpha\beta}$ is an even function of $\bd{q}$.

This ``hermitian condition" is critical for the validity of the superfluid weight formula Eq.~\ref{eq:dsformula} since it makes the last term of Eq.~\ref{eq:lastvanish} vanishes. It is an extension of the uniform diagonal case where a gauge transform is needed to bring all the diagonal entries of $\hat{\Delta}_\bd{q}$ real \cite{peotta2015superfluidity}. Of course, this condition holds at $\bd{q}=0$ because we prescribe a hermitian $\hat{\Delta}$ there; for small $\bd{q}\neq0$ state, this condition needs to be checked numerically.

\subsection{S2.3: Periodicity of Time Reversal Invariant Momentum (TRIM)}
TRS trivially imposes this periodicity since it simply says that the state of $\bd{q}$ is the same as the state of $\bd{q}+\bd{q}_0$, if $\bd{q}_0$ is a TRIM. This can be seen by a variable replacement $(\bd{k}+\bd{q}+\bd{q}_0\uparrow,-\bd{k}+\bd{q}+\bd{q}_0\downarrow)\stackrel{\bd{k}+\bd{q}_0\rightarrow\bd{k}'}{\Longrightarrow}(\bd{k}'+\bd{q}\uparrow,-\bd{k}'+\bd{q}\downarrow)$. Or from the self-consistency equations, one can check that if $\mu_\bd{q}$ and $\hat{\Delta}(\bd{k})_\bd{q}$ is a known solution for $(\bd{k},\bd{q})$, then $\mu_\bd{q}$ and $\hat{\Delta}(\bd{k}+\bd{q}_0)_\bd{q}$ is a solution for $(\bd{k},\bd{q}+\bd{q}_0)$, such that $\mu_{\bd{q}+\bd{q}_0}= \mu_\bd{q}$ and $\hat{\Delta}(\bd{k}-\bd{q}_0)_{\bd{q}+\bd{q}_0}= \hat{\Delta}(\bd{k})_\bd{q}$. As a result, $\Omega_m(\bd{q}+\bd{q}_0)=\Omega_m(\bd{q})$ and $F_m(\bd{q}+\bd{q}_0)=F_m(\bd{q})$.


\section{S3: Derivation of the Two-band Geometric Tensor Formula}
In this section, we sketch the derivation of two-band geometric tensor $G^m_{\mu\nu}(\bd{k})$ in Eq.~6 of the main text.\\

Following Eq.~\ref{eq:gexpression}, when $\hat{\Delta}$ is hermitian, we have
\be\label{eq:ghermitian}
\begin{split}
G^m_{\mu\nu}(\bd{k})=&2\bigg\{2\la u_{m,\bd{k}}|\hat{\Delta}|u_{m,\bd{k}}\ra\big[\re\la\partial_\mu u_{m,\bd{k}}|\hat{\Delta}|\partial_\nu u_{m,\bd{k}}\ra-\re\la u_{m,\bd{k}}|\hat{\Delta}|\partial_\mu\partial_\nu u_{m,\bd{k}}\ra\big]\\
&-4\im\la u_{m,\bd{k}}|\hat{\Delta}|\partial_\mu u_{m,\bd{k}}\ra\im\la u_{m,\bd{k}}|\hat{\Delta}|\partial_\nu u_{m,\bd{k}}\ra\bigg\}.
\end{split}
\ee
The Bloch function of valence (-) and conduction (+) band of the two-band model is
\be\label{eq:blochfunction}
u_{\pm,\bd{k}}=\frac{1}{\sqrt{2(1\mp\hat{h}_z)}}\begin{pmatrix}
\pm(\hat{h}_x-i\hat{h}_y)\\
1\mp\hat{h}_z
\end{pmatrix}.
\ee
Insert this and pairing matrix $\Delta_0\hat{\mathcal{I}}+\Delta_z\hat{\sigma}_z$ into Eq.~\ref{eq:ghermitian}, we find
\be
G^\pm_{\mu\nu}(\bd{k})=\frac{1}{(1+\hat{h}_z)^3}\big[\Delta_0^2A(\bd{k})+(1+\hat{h}_z)\Delta_z^2B(\bd{k})\mp\Delta_0\Delta_zC(\bd{k})\big],
\ee
where
\be
\begin{array}{c}
A(\bd{k})=2(1+\hat{h}_z)^3(\partial_\mu \hat{h}_x\partial_\nu \hat{h}_x+\partial_\mu \hat{h}_y\partial_\nu \hat{h}_y+\partial_\mu \hat{h}_z\partial_\nu \hat{h}_z),\\
\\
B(\bd{k})=-2(1+\hat{h}_z)^2(\partial_\mu \hat{h}_x\partial_\nu \hat{h}_x+\partial_\mu \hat{h}_y\partial_\nu \hat{h}_y+\hat{h}_z\partial_\mu\partial_\nu \hat{h}_z),\\
\\
C(\bd{k})=2(1+\hat{h}_z)^3\partial_\mu\partial_\nu \hat{h}_z.
\end{array}
\ee
Therefore
\be
\begin{split}
\label{eq:gtensorsupp}
G^\pm_{\mu\nu}(\bd{k})=&2\Delta_0^2(\partial_\mu \hat{h}_x\partial_\nu \hat{h}_x+\partial_\mu \hat{h}_y\partial_\nu \hat{h}_y+\partial_\mu \hat{h}_z\partial_\nu\hat{h}_z)\\
&-2\Delta_z^2(\partial_\mu \hat{h}_x\partial_\nu \hat{h}_x+\partial_\mu \hat{h}_y\partial_\nu \hat{h}_y+\hat{h}_z\partial_\mu\partial_\nu\hat{h}_z)\\
&\mp2\Delta_0\Delta_z\partial_\mu\partial_\nu\hat{h}_z.
\end{split}
\ee

The expression for $G^\pm_{\mu\nu}(\bd{k})$ for the respective order parameter matrices, $\hat{\Delta} = \Delta_0\hat{\mathcal{I}} + \Delta_x \hat{\sigma}_x$, and $\hat{\Delta} = \Delta_0\hat{\mathcal{I}} + \Delta_y \hat{\sigma}_y$, can be recovered by sending $h_z \leftrightarrow h_x $, and $h_z \leftrightarrow h_y $ in Eq.~\label{eq:gtensorsupp} above. It is also then easy to generalize $G^\pm_{\mu\nu}(\bd{k})$ for a general Hermitian $\hat{\Delta}$.


\section{S4: Universality of the BCS-PDW Transition}
This section discusses the universality of BCS-PDW transition in multi-orbital superconductors. This transition is marked by a NPD $D^{m, geo}_{s,\mu \nu}$, which generally occurs when the NPD contributions to $G^{m}_{\mu\nu} (\bd{k})$ in the BZ become greater than the PD contributions. There are two possible scenarios i) $D^{s, geo}_{\mu \nu}$ remains regular over the whole BZ, in which case transition occurs when regions of negative contributions dominate, ii) $D^{s, geo}_{\mu \nu}$ has singular behavior at either isolated nodal points or nodal arcs in the BZ which are enclosed in the NPD $G^{m}_{\mu\nu} (\bd{k})$ region, in which case the transition occurs sharply.

In S4.1, we restrict ourselves to the case of specific two-band models and determine when $G^{m}_{\mu \nu}$ becomes NPD by analyzing Eq.~6 of the main text. In this case, for a regular $D^{m, geo}_{s,\mu \nu}$, we establish when the NPD geometric contributions overcome the PD contributions by requiring the band to be ``quasi-flat" and interaction to be ``relatively strong". In S4.2, we discuss the general case, where nodal singularities facilitate the transitions, even in a dispersive band.
\subsection{S4.1: Transition in a Two-band Hamiltonian}
Here, within the band-projection formalism, we show that $G^{m}_{\mu\nu} (\bd{k}) $ can turn NPD easily for a certain class of two-band models. We assume that the two-band model Hamiltonian
has the form $h^\uparrow(\bd{k})=\bd{h}(\bd{k})\cdot\bds{\sigma}$, with $\bd{h}(\bd{k})=(h_x(\bd{k}),h_y(\bd{k}),M)$ satisfying the condition $M\gg|h_x|,|h_y|$. This means that the band gap is almost a constant $2M$. We assume that this band gap does not change much either in some regions of the BZ or over the whole BZ, as in the case of a flat band. We show that for the pairing matrix $\hat{\Delta} = \Delta_0\hat{\mathcal{I}}+\Delta_z\hat{\sigma}_z$ under certain conditions, $D^{geo}_{s,\mu\nu}$ will always turn NPD. We calculate $G_{\mu \nu}$ for the valence band; the conduction band gives the same results.

For the projection to be valid, we require $\Delta_0,\Delta_z\ll M$. We first look at the geometric term in Eq.~5 of the main text. Now $\hat{\bd{h}}=(h_x,h_y,M)/\sqrt{h_x^2+h_y^2+M^2}$ is a unit vector. By assumption $M\gg|h_x|,|h_y|$, one can Taylor expand Eq.~6 in orders of $M$, yielding
\be\label{eq:gquasiflat}
\begin{split}
G^v_{\mu\nu}(\bd{k})=&\frac{2\Delta_0^2}{M^2}(\partial_\mu h_x\partial_\nu h_x+\partial_\mu h_y\partial_\nu h_y)-\frac{\Delta_0\Delta_z}{M^2}\partial_\mu\partial_\nu(h_x^2+h_y^2)\\
&-\frac{2\Delta_z^2}{M^2}\big[\partial_\mu h_x\partial_\nu h_x+\partial_\mu h_y\partial_\nu h_y-\frac{1}{2}\partial_\mu\partial_\nu(h_x^2+h_y^2)\big]+O\big(\frac{1}{M^4}\big).
\end{split}
\ee
To get Eq.~\ref{eq:gquasiflat}, we dropped the $\Delta_0^2\partial_\mu\hat{h}_z\partial_\nu\hat{h}_z$ term since it contributes $O(1/M^4)$. Here, the terms $\partial_\mu h_x\partial_\nu h_x$ and $\partial_\mu h_y\partial_\nu h_y$ are positive semidefinite, since $\partial_\mu h_x\partial_\nu h_x$ has the matrix form 
\be 
\begin{pmatrix}a^2&ab\\ab&b^2\end{pmatrix},
\ee and the bilinear form 
\be 
\bd{x}^T
\begin{pmatrix}a^2&ab\\ab&b^2\end{pmatrix}
\bd{x} \geq 0.
\ee
The term $\partial_\mu\partial_\nu(h_x^2+h_y^2)$ is instead indefinite. 

However, one notices that $\partial_\mu\partial_\nu(h_x^2+h_y^2)$ is related to the flatness of the band. For a perfectly flat band, $h_x^2+h_y^2$ is a constant, and $\partial_\mu\partial_\nu(h_x^2+h_y^2)$ vanishes; for a ``quasi-flat" band, which is \tit{defined} by condition $|\partial_\mu\partial_\nu(h_x^2+h_y^2)|\ll|\partial_\mu h_x\partial_\nu h_x|,|\partial_\mu h_y\partial_\nu h_y|$, therefore, it can be dropped, so we are left with
\be\label{eq:geometry2band}
G^v_{\mu\nu}(\bd{k})\approx\frac{2(\Delta_0^2-\Delta_z^2)}{M^2}(\partial_\mu h_x\partial_\nu h_x+\partial_\mu h_y\partial_\nu h_y),
\ee
which is positive (negative) definite for $|\Delta_z/\Delta_0|>1$ ($<1$). In the case of flat bands, where the quasi-flat band condition is presumed to be valid over the whole BZ, the BCS-PDW transition occurs when $|\Delta_z/\Delta_0|>1$.

One can further characterize contribution of the ``conventional term"  given by Eq.~4 in the ``quasi-flat" band limit,
\be
-\xi_{m,\bd{k}}\partial_\mu\partial_\nu\xi_{m,\bd{k}}=-\frac{\xi_{m,\bd{k}}}{2M}\partial_\mu\partial_\nu(h_x^2+h_y^2)+O\big(\frac{1}{M^3}\big).
\ee
In additional to the ``quasi-flat" condition, if we also require $|\delta\xi_{m,\bd{k}}/M|\ll|\Delta^2/M^2|$ (or $\Delta\gg\sqrt{M\delta\xi_{m,\bd{k}}}$), then the geometric term will dominate. Here $\Delta$ is of the order of $max\{\Delta_0,\Delta_z\}$ (we assume $\Delta_0\neq\Delta_z$ so the geometric term Eq.~\ref{eq:geometry2band} does not cancel); $\delta\xi_{m,\bd{k}}$ is the bandwidth. In this case, the transition occurs when $|\Delta_z/\Delta_0|>1$.

In the presence of nodal zeroes in the quasiparticle spectrum, $E_{m,\bd{k}} \to 0$ at some point ${\bd{k}_0}$ or neighborhood of ${\bd{k}_0}$. Depending on the behavior of $E_{m,\bd{k}} \to 0$, when they are enclosed in the NPD regions of $G^m_{\mu \nu}$, $D^{m, geo}_{s,\mu\nu}$ is no longer regular and can diverge,  the transition occurs sharply at $|\Delta_z/\Delta_0| \geq 1$. This can happen for the two-band model discussed above, even when the quasi-flat condition is only valid over a small region of the BZ that includes nodal zeroes. In 2D, there are two possibilities for nodal zeroes: i) isolated nodal points or ii) nodal arcs. Whether $D^{m,geo}_{s,\mu \nu}$ is singular depends on the behavior of $E_{m,\bd{k}}$ as it approaches the nodal zeroes. We analyze this for the general case in the next section. 

\subsection{S4.2: Transition Driven by Nodal Singularities}
When a band has dispersion, both the ``quasi-flat" and the ``strong interaction" condition may fail. Now, we discuss another scenario when the BCS-PDW transition can happen, regardless of dispersions, when the quasiparticle spectrum has nodal zeroes. This result is not limited to two-band models. Here by ``quasiparticle spectrum", we mean the BCS $\bd{q}=0$ spectrum, $E_{m,\bd{k}}$, since the superfluid weight formula Eq.~4 and 5 only depend on this spectrum.

We will analyze the nodal points and nodal arcs separately since they have different analytical properties. From Eq.~\ref{eq:dsexpression2}, the conventional term can only give regular contributions, even near nodal zeroes. This is because as $E_{m,\bd{k}}=\sqrt{\xi_{m,\bd{k}}^2+|\Delta_{m,\bd{k}}(0)|^2}\rightarrow0$, we have $|\xi_{m,\bd{k}}|<E_{m,\bd{k}}$, and $\partial_\mu\partial_\nu\xi_{m,\bd{k}}$ is a regular function, so the conventional term
\be
-\int\frac{dk^2}{(2\pi)^2}\frac{\xi_{m,\bd{k}}}{E_{m,\bd{k}}}\partial_\mu\partial_\nu\xi_{m,\bd{k}}
\ee
is always finite. Therefore, if $D^{m,geo}_{s,\mu\nu}$ exhibits a negative divergence due to nodal singularities, the conventional terms will be subdominant and can be ignored.  

Now we look at the geometric term. If the quasiparticle spectrum $E_{m,\bd{k}}$ contains an isolated nodal point $\bd{k}_0$, which for simplicity is assumed not to coincide with any zeroes of $G^m_{\mu\nu}(\bd{k})$ (for BHZ and two-band models, we will see counter-examples to this assumption in S5), then at a small neighborhood of $\bd{k}_0$, $G^m_{\mu\nu}(\bd{k})$ can be taken as constant. Let's assume circular symmetry for $E_{m,\bd{k}}$ near $\bd{k}_0$. For the case of linear dispersion, we have $E_{m,\bd{k}}\sim |\bd{k}-\bd{k}_0|$. Then the geometric contribution from the nodal point is
\be
\int_{\text{near}\,\bd{k}_0}\frac{dk^2}{(2\pi)^2}\frac{G^m_{\mu\nu}(\bd{k})}{2E_{m,\bd{k}}}=c\int dk^2\frac{1}{|\bd{k}-\bd{k}_0|}=c\int_0^\kappa 2\pi kdk\frac{1}{|\bd{k}-\bd{k}_0|}=2\pi c\kappa,
\ee
where $\kappa$ is a small cutoff radius and $c$ is a constant. This says that a nodal point with dispersion $E_{m,\bd{k}}\sim |\bd{k}-\bd{k}_0|$ always gives a regular contribution to the superfluid weight.

Similarly, for a nodal point with $E_{m,\bd{k}}\sim |\bd{k}-\bd{k}_0|^2$, it will give logarithm divergence. We, therefore, conclude that \tit{for 2D materials, if there are nodal points in the $\bd{q}=0$ quasiparticle spectrum with $E_{m,\bd{k}}\sim |\bd{k}-\bd{k}_0|^\alpha$ ($\alpha\geq2$), and if all these nodal points fall into the NPD region of $G^m_{\mu\nu}(\bd{k})$, then it is guaranteed that $D^m_{s,\mu\nu}$ is NPD and the BCS $\bd{q}=0$ state is unstable}.

Next, we look at nodal arcs. Nodal arcs are defects with one more dimension than nodal points, making it easier to give singular contributions to $D^m_{s,\mu\nu}$. For the geometric contribution, if we assume a linear dispersion $E_{m,\bd{k}}\sim k$, with $k$ the distance between $\bd{k}$ point and the nodal arc $L$, then we have
\be\label{eq:nodalline}
\int_{\text{near}\,L}\frac{dk^2}{(2\pi)^2}\frac{G^m_{\mu\nu}(\bd{k})}{2E_{m,\bd{k}}}=c\int dk^2\frac{1}{k}=cdl\int_0^\kappa dk\frac{1}{k},
\ee
where $dl$ is the length of the nodal arc segment. The logarithm divergence of Eq.~\ref{eq:nodalline} tells us that \tit{for 2D materials, if there are nodal arcs in the $\bd{q}=0$ quasiparticle spectrum with $E_{m,\bd{k}}\sim k^\alpha$ ($\alpha\geq1$), and if all these nodal arcs are enclosed in the NPD region of $G^m_{\mu\nu}(\bd{k})$, then it is guaranteed that $D^m_{s,\mu\nu}$ is NPD and the BCS $\bd{q}=0$ state is unstable}.

Last, we want to highlight that nodal arcs or circles are unusual in 2D materials. From $E_{m,\bd{k}}=\sqrt{\xi_{m,\bd{k}}^2+|\Delta_{m,\bd{k}}(0)|^2}$, the nodality condition is to satisfy
\be
\varepsilon_{m,\bd{k}}-\mu=0,\,\,\,\text{and}\,\,\,\Delta_{m,\bd{k}}(0)=\la u_{m,\bd{k}}|\hat{\Delta}(\bd{k})|u_{m,\bd{k}}\ra=0
\ee
simultaneously, which are two constraints for two variables $k_x,k_y$. Usually, this leads to isolated nodal points. To have nodal arcs, the gap function $\Delta_{m,\bd{k}}(0)$ and dispersion $\varepsilon_{m,\bd{k}}$ must share certain symmetries, such that the lines individually determined by the two equations above coincide. A trivial case is when the band is perfectly flat. When $\mu$ is aligned with the flat band, only one constraint $\Delta_{m,\bd{k}}(0)=0$ is left, so nodal arcs can show up easily.

When there are nodal singularities present, we give the integral evaluation below for $E_{m,\bd{k}}\sim k^\alpha$, by including a broadening constant $\epsilon$. The broadening can come from either the imperfection of materials or misaligning of the chemical potential, such that the quasiparticle spectrum has a small gap $2\epsilon$.
\be
\begin{array}{ll}
\text{Nodal point } E_{m,\bd{k}}\sim |\bd{k}-\bd{k}_0|^2 \text{ or nodal arc } E_{m,\bd{k}}\sim k:&\int dk\frac{1}{\sqrt{k^2+\epsilon^2}}\sim -\ln\epsilon,\\
\\
\text{Nodal point } E_{m,\bd{k}}\sim |\bd{k}-\bd{k}_0|^3 \text{ or nodal arc } E_{m,\bd{k}}\sim k^2:&\int dk\frac{1}{\sqrt{k^4+\epsilon^2}}\sim \epsilon^{-1/2},...
\end{array}
\ee
%


\section{S5: Numerical Calculation of the $D^v_{s,\mu\nu}$ Instability Curve}
This section provides details for calculating the $D^v_{s,\mu\nu}$ instability curve. In S5.1, we discuss the center symmetry of the phase diagram of the BHZ model, which is why only the $\Delta_z/\Delta_0>0$ half of the diagram is given in Fig.~3 of the main text. In S5.2, we discuss the nodal points of the flattened BHZ model, which are regular and can not drive the BCS-PDW transition. In S5.3, we discuss the nodal circles, which give singular contributions and drive the transition. In S5.4, we discuss the large-mass atomic limit of the model. The nodality discussion in S5.2 and S5.3 apply to generic topological two-band systems, so is not limited to the flattened BHZ model.
\subsection{S5.1: Centrosymmetry of the Phase Diagram}
The $D^v_{s,\mu\nu}$ instability curve is determined from condition $\text{det}D^v_{s,\mu\nu}=0$ and $\tr{D^v_{s,\mu\nu}}\ge0$. The curve coincides with the BCS-PDW phase boundary when the transition is second-order. For the BHZ model with $\Delta_0\hat{\mathcal{I}}+\Delta_z\hat{\sigma}_z$ type pairing, $C_4$ rotation symmetry is preserved, so the condition above is equivalent to $D^v_s\equiv D^v_{s,xx}=D^v_{s,yy}=0$.

The instability curve on the $\Delta_z/\Delta_0<0$ side is centrosymmetric to the $\Delta_z/\Delta_0>0$ side about $(m_0,\Delta_z/\Delta_0)=(0,0)$, due to the symmetry between parameters $(m_0,\Delta_z/\Delta_0,k_x,k_y)$ and $(-m_0,-\Delta_z/\Delta_0,k_x+\pi,k_y+\pi)$ for the flattened BHZ model:
\be\label{eq:sym1}
\begin{split}
&G^v_{xx}(m_0,\Delta_z/\Delta_0,k_x,k_y)=G^v_{xx}(-m_0,-\Delta_z/\Delta_0,k_x+\pi,k_y+\pi),\\
&E_{v,\bd{k}}(m_0,\Delta_z/\Delta_0,k_x,k_y)=E_{v,\bd{k}}(-m_0,-\Delta_z/\Delta_0,k_x+\pi,k_y+\pi).
\end{split}
\ee
This leads to $D^v_{s}(m_0,\Delta_z/\Delta_0)=D^v_{s}(-m_0,-\Delta_z/\Delta_0)$, therefore we only focus on the $\Delta_z/\Delta_0>0$ side.
\subsection{S5.2: Nodal Points}
\begin{figure}[h]
\captionsetup{singlelinecheck = false, justification=raggedright}
\centering
\begin{subfigure}{0.24\textwidth}
	\centering
	\includegraphics[width=0.98\textwidth]{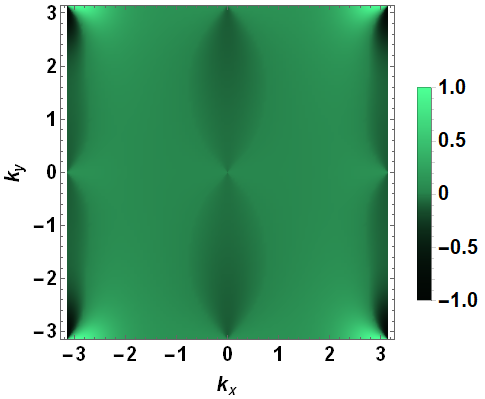}
	\caption{\centering}
\end{subfigure}
\begin{subfigure}{0.24\textwidth}
	\centering
	\includegraphics[width=0.95\textwidth]{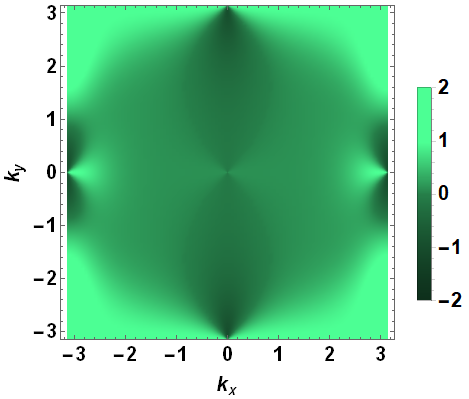}
	\caption{\centering}
\end{subfigure}
\begin{subfigure}{0.24\textwidth}
	\centering
	\includegraphics[width=0.98\textwidth]{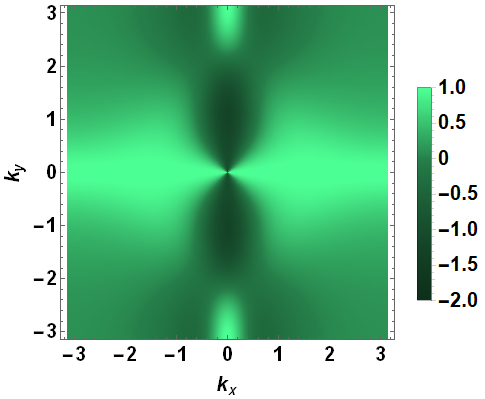}
	\caption{\centering}
\end{subfigure}
\begin{subfigure}{0.24\textwidth}
	\centering
	\includegraphics[width=0.98\textwidth]{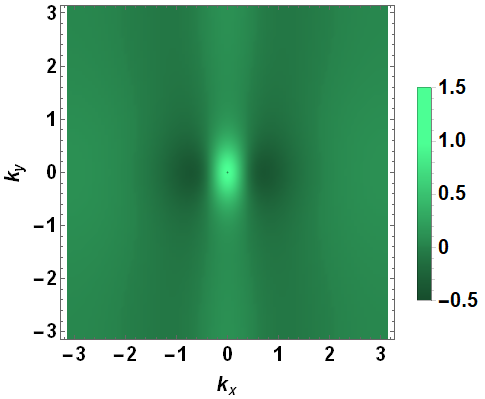}
	\caption{\centering}
\end{subfigure}
	\caption{Density plot of $G^v_{xx}(\bd{k})/2E_{v,\bd{k}}$ at $\Delta_z/\Delta_0=1$ with nodal points. $G^v_{xx}(\bd{k})<0$ region is slightly shaded. Due to the cancellation of zeroes of $G^v_{xx}(\bd{k})$, $G^v_{xx}(\bd{k})/2E_{v,\bd{k}}$ are finite at the nodal points. (a) $m_0=3$, four nodal points at $\Gamma,X,X',M$; (b) $m_0=1$, three nodal points at $\Gamma,X,X'$; (c) $m_0=-1$, one nodal point at $\Gamma$; (d) $m_0=-3$, no nodal points.}
	\label{fig:nodalpt}
\end{figure}
When the chemical potential is aligned with the flat valence band, the $\bd{q}=0$ quasiparticle energy is
\be\label{eq:quasievk}
E_{v,\bd{k}}=|\Delta_{v,\bd{k}}(0)|=|\Delta_0-\Delta_z\hat{h}_z(\bd{k})|.
\ee
One notices that nodal zeroes can appear in the spectrum only when $\Delta_z/\Delta_0\ge1$. At $\Delta_z/\Delta_0=1$, the only possible nodal points are the four TRIM $\Gamma(0,0),X(\pi,0),X'(0,\pi),M(\pi,\pi)$, where $\hat{h}_x=\hat{h}_y=0$ and $\hat{h}_z=\sgn(m_0+\cos k_x+\cos k_y)$. We tabulate the conditions that they become nodal points in Table~\ref{tab:nodalcondition}.
\begin{table}[h]
\centering
\begin{tabular}{c|c|c}
    & condition & which phases\\
    \hline
    $\Gamma(0,0)$ & $\sgn(m_0+2)=1$ & $m_0>-2$ \\
    \hline
    $X(\pi,0)$ & $\sgn(m_0)=1$ & $m_0>0$ \\
    \hline
    $X'(0,\pi)$ & $\sgn(m_0)=1$ & $m_0>0$ \\
    \hline
    $M(\pi,\pi)$ & $\sgn(m_0-2)=1$ & $m_0>2$ \\
    \hline
\end{tabular}
\caption{Nodality condition for the four TRIM at $\Delta_z/\Delta_0=1$.}
\label{tab:nodalcondition}
\end{table}

Therefore at $\Delta_z/\Delta_0=1$, for the trivial BHZ phase $m_0>2$, all the four TRIM are nodal points; for $0<m_0<2$, $\Gamma,X,X'$ are nodal points; for $-2<m_0<0$, $\Gamma$ is the only nodal point; for $m_0<-2$, no nodal points. One can Taylor expand Eq.~\ref{eq:quasievk} to show that these nodal points are all \tit{quadratic}, e.g. at $\Gamma$ point
\be
E_{v,\bd{k}}\approx\frac{1}{2(m_0+2)^2}(k_x^2+k_y^2)+O(k^3),\,\,\,\text{ for }m_0>-2.
\ee
However, these nodal points all coincide with zeroes of $G^v_{xx}(\bd{k})$ (which are also the intersection point of positive and negative $G^v_{xx}(\bd{k})$ regions), such that $G^v_{xx}(\bd{k})/2E_{v,\bd{k}}$ remains finite (see Fig.~\ref{fig:nodalpt}). As a result, $D_s^v$ is positive at $\Delta_z/\Delta_0=1$ for all BHZ phases.
\subsection{S5.3: Nodal Circles}
\begin{figure}[h]
\captionsetup{singlelinecheck = false, justification=raggedright}
\centering
\begin{subfigure}{0.24\textwidth}
	\centering
	\includegraphics[width=0.98\textwidth]{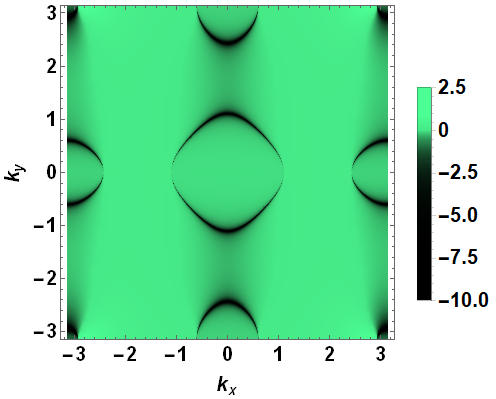}
	\caption{\centering}
\end{subfigure}
\begin{subfigure}{0.24\textwidth}
	\centering
	\includegraphics[width=0.98\textwidth]{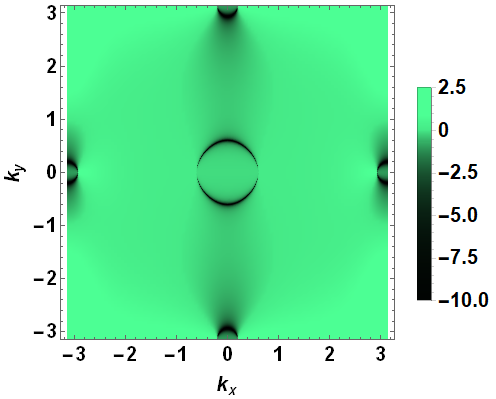}
	\caption{\centering}
\end{subfigure}
\begin{subfigure}{0.24\textwidth}
	\centering
	\includegraphics[width=0.98\textwidth]{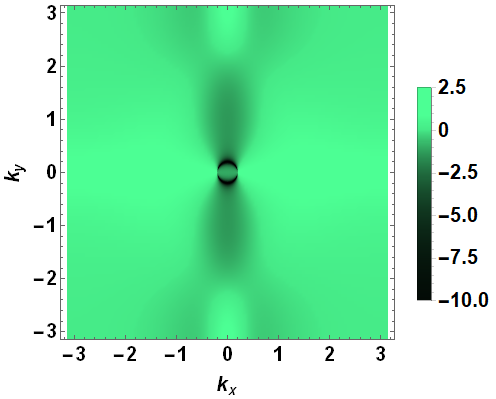}
	\caption{\centering}
\end{subfigure}
\begin{subfigure}{0.24\textwidth}
	\centering
	\includegraphics[width=0.98\textwidth]{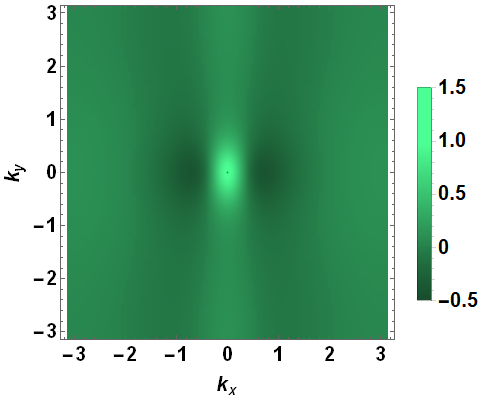}
	\caption{\centering}
\end{subfigure}
	\caption{Density plot of $G^v_{xx}(\bd{k})/2E_{v,\bd{k}}$ at $\Delta_z/\Delta_0=1.02$ with nodal circles. $G^v_{xx}(\bd{k})<0$ region is slightly shaded, and the nodal circles are shown in black, which diverges to $-\infty$. (a) $m_0=3$, four nodal circles centered at $\Gamma,X,X',M$; (b) $m_0=1$, three nodal circles at $\Gamma,X,X'$; (c) $m_0=-1$, one nodal circle at $\Gamma$; (d) $m_0=-3$, no nodal circles.}
	\label{fig:nodalcircle}
\end{figure}
For the $m_0>-2$ phases of the flattened BHZ model, it turns out that $\Delta_z/\Delta_0=1$ is a critical point. When $\Delta_z/\Delta_0=1+\eta_+$, with $\eta_+$ an infinitesimal positive value, these nodal points expand into nodal circles, which are always enclosed in the negative $G^v_{xx}(\bd{k})$ regions (Fig.~\ref{fig:nodalcircle}). These nodal circles all have \tit{linear} dispersion. e.g. for the one at $\Gamma$ point, one can do an expansion of $E_{v,\bd{k}}$ near some zero on the $k_x$ axis, $(k_0,0)$:
\be
E_{v,\bd{k}}\approx\Delta_z\frac{|\sin k_0(m\cos k_0+\cos k_0+1)|}{[(m+\cos k_0+1)^2+\sin^2 k_0]^{3/2}}|k_x-k_0|+O((k_x-k_0)^2),\,\,\,\text{ for }m_0>-2.
\ee
Then according to our discussion in S4.2, these nodal circles give a negative singular contribution to $D^v_s$, from which we conclude that the BCS-PDW transition happens at $\Delta_z/\Delta_0=1+\eta_+$. On the other hand, the $m_0<-2$ phase has no nodal point or circles, so the transition occurs at a higher $\Delta_z/\Delta_0$ value.

The fact that nodal circles for a generic two-band model are enclosed in the NPD $G^v_{\mu\nu}$ regions can be understood as follows. Assuming a two-band hamiltonian which has the maximum value of $\hat{h}_z$ to be 1 at some point $\bd{k}_0$, then at $\Delta_z/\Delta_0=1+\eta$, with $\eta$ very small, $\hat{\bd{h}}(\bd{k})$ can be expanded as $\hat{\bd{h}}(\bd{k})\approx(vp_x,vp_y,1-\frac{v^2}{2}(p_x^2+p_y^2))$ at the neighborhood of nodal point ($\bd{p}=\bd{k}-\bd{k}_0$ is the small momentum shift measured from the nodal point $\bd{k}_0$). Here $v$ is an expansion coefficient, and we have assumed circular symmetry around the nodal point. By solving the equation for nodal zeroes, $\Delta_0-\Delta_z\hat{h}_z(\bd{k})=0$, we find
\be\label{eq:radius}
p^2=\frac{2\eta}{v^2}+O(\eta^2).
\ee
For $\eta<0$, it has no solution; for $\eta\geq0$, the nodal circle has radius $(1/v)\sqrt{2\eta}$ to the lowest order of $\eta$. On the other hand, if we plug $\hat{\bd{h}}$ into the $G^v_{\mu\nu}(\bd{k})$ expression Eq.~6, we find 
\be
G^v_{xx}(\bd{k})=\Delta_0^2\big[-2\eta v^2+(1-2\eta)v^4p_x^2-(1+2\eta)v^4p_y^2\big]+O(\eta^2).
\ee
Therefore to the lowest order of $\eta>0$, the negative $G^v_{xx}(\bd{k})$ region is given by
\be
p_x^2\leq\frac{2\eta}{v^2}+(1+4\eta)p_y^2+O(\eta^2),
\ee
which always encloses and is \tit{tangent} to the nodal circle in equation Eq.~\ref{eq:radius}. This is exactly what we observe in Fig.~\ref{fig:nodalcircle}. At $\Delta_z/\Delta_0=1$ or $\eta=0$, we have $G^v_{xx}(\bd{k})\approx \Delta_0^2v^4(p_x^2-p_y^2)$, which is why we always have the nodal points coinciding with zeroes of $G^v_{xx}(\bd{k})$ and sitting in the intersection point of positive and negative regions, as shown in Fig.~\ref{fig:nodalpt}. These results are also schematically shown in Fig.~1 of the main text.

For a more interesting $D^v_s$ instability curve, we get rid of the singularity of nodal circles by adding a broadening constant $\epsilon$ to the quasiparticle energy, $E_{v,\bd{k}}\rightarrow\sqrt{\epsilon^2+[\Delta_0-\Delta_z\hat{h}_z(\bd{k})]^2}$, which gives Fig.~2(a) in the main text. Alternatively, one can also add weak dispersion to the flat band. In Fig.~2(b), we use the actual BHZ model valence band energy to model the weak dispersion, which is
\be
\delta\epsilon_\bd{k}=-\sqrt{t^2\sin^2 k_x+t^2\sin^2 k_y+(M_0+t\cos k_x+t\cos k_y)^2}.
\ee
$t$ is the hopping parameter, and $M_0$ is the unscaled mass of the BHZ model (to distinguish from the scaled mass $m_0$ in the main text). For weak dispersion, we set $t=0.01\Delta_0$ and $M_0/t=m_0$ (which are $\pm3,\pm1$ for the four phases). In the $D_s$ expression Eq.~4 and 5, we replace the kinetic energy everywhere with $\delta\xi_\bd{k}=\delta\epsilon_\bd{k}-\mu$, while keeping all the geometric quantities ($G^v_{\mu\nu}$, $\hat{h}_z$, etc.) unchanged. The chemical potential $\mu$ is put at the middle of the band, e.g., for $m_0=3$ phase, we have $t=0.01$, $M_0=0.03$, and $-0.05\leq\delta\epsilon_\bd{k}\leq-0.01$, so we set $\mu=-0.03$. The effect of weak dispersion turns out to be similar to broadening. Due to broadening or weak dispersion, the instability curve is shifted up in the $m_0>-2$ phases but is unchanged for the $m_0<-2$ phase since there are no nodal circles there.
\subsection{S5.4: Large-mass Atomic Limit}
We look at the limit $m_0\rightarrow\pm\infty$ of the BHZ phase diagram. This is an atomic limit that is slightly different from the ``quasi-flat" band discussion in S4.1. Still, we will get a similar result---the $D^v_s$ instability curve asymptotically goes to $\Delta_z/\Delta_0=1$ as $m_0\rightarrow\pm\infty$, which is shown in Fig.~3 of the main text. This limiting behavior is independent of the broadening $\epsilon$.

One notices that although the $m_0>2$ phase has nodal zeroes, it is identical to the $m_0<-2$ phase if $|m_0|$ becomes large. The reason is that as $m_0\rightarrow+\infty$, we find $\hat{h}_z(\bd{k})\rightarrow 1$, $\forall\bd{k}\in $BZ, which means the nodal zeroes only show up in an infinitesimal range $1\leq\Delta_z/\Delta_0<1+\eta_+$. i.e. effectively the $m_0\rightarrow+\infty$ limit has \tit{no} nodal zeroes and $E_{v,\bd{k}}\rightarrow|\Delta_0-\Delta_z|$, which is a nonzero constant when $\Delta_z/\Delta_0$ is slightly greater than 1. If there is broadening, this constant is modified to $\sqrt{\epsilon^2+(\Delta_0-\Delta_z)^2}$. Similarly, for $m_0\rightarrow-\infty$, we have $\hat{h}_z(\bd{k})\rightarrow -1$ and $E_{v,\bd{k}}\rightarrow\sqrt{\epsilon^2+(\Delta_0+\Delta_z)^2}$.

Taylor expansion of $G^v_{xx}(\bd{k})$ around $m_0=\pm\infty$ gives
\be
G^v_{xx}(\bd{k})\approx \frac{\Delta_0^2}{m_0^2}[\cos(2k_x)+1]+\frac{\Delta_z^2}{m_0^2}[\cos(2k_x)-1]-2\Delta_0\Delta_z\frac{|m_0|}{m_0^3}\cos(2k_x)+O(\frac{1}{m_0^3}).
\ee
In the integral $D^v_{s}=\int d^2kG^v_{xx}(\bd{k})/2E_{v,\bd{k}}$, $E_{v,\bd{k}}$ is almost a constant, and $\int d^2k\cos(2k_x)=0$, therefore
\be
D^v_{s}\approx\frac{\Delta_0^2-\Delta_z^2}{m_0^2}\int d^2k\frac{1}{2E_{v,\bd{k}}},
\ee
which changes sign as $|\Delta_z/\Delta_0|$ passes 1.
\section{S6: Numerical Calculation of the Free Energy $F_v(\bd{q})$}
In this section, we provide details for the numerical calculation of band-projected grand potential $\Omega_v(\bd{q})$ and free energy $F_v(\bd{q})$, for the flattened BHZ model with $\Delta_0\hat{\mathcal{I}}+\Delta_z\hat{\sigma}_z$ type pairing. After the valence band projection, let's choose the valence band as the energy reference point by setting $\varepsilon_{v,\bd{k}}=0$. Assuming an onsite interaction $V^{(\alpha\beta,\gamma\delta)}_{\bd{k}\bd{k}'}=-(1/N)U_{\alpha\beta}\delta_{\alpha\gamma}\delta_{\beta\delta}$, then the self-consistency equation Eq.~\ref{eq:deltaperiod1}, \ref{eq:muperiod1} for the valence band become
\be\label{eq:finaldeltaself}
\hat{\Delta}_{\bd{q},\alpha\beta}=\frac{U_{\alpha\beta}}{2}\int\frac{d^2k}{(2\pi)^2}u_{v\beta}(\bd{k}-\bd{q})^*u_{v\alpha}(\bd{k}+\bd{q})\frac{\Delta_{v,\bd{k}}(\bd{q})}{\sqrt{\mu_\bd{q}^2+|\Delta_{v,\bd{k}}(\bd{q})|^2}},
\ee
\be\label{eq:finalmuself}
\int\frac{d^2k}{(2\pi)^2}\frac{\mu_\bd{q}}{\sqrt{\mu_\bd{q}^2+|\Delta_{v,\bd{k}}(\bd{q})|^2}}=\frac{N_{e,v}}{N}-1.
\ee

By controlling the electron density, one can set $\mu$ for the $\bd{q}=0$ state. To cure the divergence due to nodal singularities, we turn on a small broadening by setting $\mu_{\bd{q}=0}=\epsilon$, and solve both $\mu_\bd{q}$ and $\hat{\Delta}_\bd{q}$ self-consistently. We choose $\hat{\Delta}=\Delta_0\hat{\mathcal{I}}+\Delta_z\hat{\sigma}_z$ at $\bd{q}=0$, with $\Delta_0,\Delta_z$ both real. Then $U_{\alpha\beta}$ are $U_{11}=U_1$, $U_{22}=U_2$, and 0 otherwise, according to Eq.~\ref{eq:finaldeltaself}. Note that even at $\Delta_z=0$, it is possible that $U_1\neq U_2$ because the two orbitals may have unequal weight in the valence band. For $\bd{q}\neq0$, $\mu_\bd{q},\hat{\Delta}_\bd{q}$ are solved from Eq.~\ref{eq:finaldeltaself}, Eq.~\ref{eq:finalmuself} using Newton-Raphson iteration method. The solution $\mu_\bd{q}$ and $\hat{\Delta}_\bd{q}$ are required to be continuous in $\bd{q}$ space and smoothly connected to $\epsilon$ and $\Delta_0\hat{\mathcal{I}}+\Delta_z\hat{\sigma}_z$ at $\bd{q}=0$. Numerically we find that at any $\bd{q}$, $\hat{\Delta}(\bd{q})$ is also a real matrix, of the form $\text{diag}(\Delta_{11}(\bd{q}),\Delta_{22}(\bd{q}))$, so the ``hermitian condition" stated in S2.2 is satisfied. Then we use Eq.~\ref{eq:grandpotential2} and $F_v(\bd{q})=\Omega_v(\bd{q})+\mu_\bd{q}N_{e,v}$ to calculate $\Omega_v(\bd{q})$ and $F_v(\bd{q})$.
\section{S7: Phase Diagrams in the $U_{\alpha\beta}$ Parameter Space}
In this section we provide the BCS-PDW phase diagrams of the flattened BHZ model in $U_{\alpha\beta}$ space for various BHZ phases.\\

By the convention of Eq. 1 in the maintext, $U_{\alpha\beta}>0$ means attractive interactions. Parameter $\hat{\Delta}_{11}=\Delta_0+\Delta_z,\hat{\Delta}_{22}=\Delta_0-\Delta_z$ and $U_{11},U_{22}$ are related through self-consistency equation Eq. \ref{eq:finaldeltaself} by setting $\bd{q}=0$. We consider the simplest case when the chemical potential is very close to the flat band, i.e. setting $\mu\ll \Delta$ in Eq. \ref{eq:finaldeltaself}. This corresponds to nearly half filling in the valence band, $n_{e,v}=N_{e,v}/N\approx 1$ ($n_{e,v}=2$ means completely filled). Then $U_{11},U_{22}$ are the only energy scales of the problem and one would find $\hat{\Delta}_{\alpha\beta}\propto U_{\alpha\beta}$, so the phase boundary in the $U$-space will be represented by rays from the origin.

After projection to the valence band, by Eq. \ref{eq:blochfunction},
\be
u_{-,\bd{k}}=\frac{1}{\sqrt{2(1+\hat{h}_z)}}\begin{pmatrix}
-(\hat{h}_x-i\hat{h}_y)\\1+\hat{h}_z
\end{pmatrix}
\ee
with
\be
(\hat{h}_x,\hat{h}_y,\hat{h}_z)=\frac{(\sin k_x,\sin k_y,m_0+\cos k_x+\cos k_y)}{\sqrt{\sin^2 k_x+\sin^2 k_y+(m_0+\cos k_x+\cos k_y)^2}}.
\ee
In the atomic limit $m_0\rightarrow\pm\infty$, $\hat{h}_z\rightarrow\pm1$ so the valence band is strongly polarized to orbital 2 or orbital 1. In the opposite limit $m_0\rightarrow0$, the two orbitals have equal weight in the band. 

The $U$-space diagram gets some new features from the polarization effect. For example, at $m_0=3$, it is strongly polarized to orbital 2. As a result, a large domain in $\Delta_z/\Delta_0$ space at $\Delta_z/\Delta_0>0$ side correspond to a small-angle domain in $U$-space, with $|U_{22}/U_{11}|\approx 0$. Since no matter in which space the total measure should be equal. As a complement to this, there is also some small range in $\Delta_z/\Delta_0$ space that corresponds to a large domain in $U$-space (e.g. in Fig. \ref{fig:udiagram1}(a), $m_0=3$, regime $1.054<\Delta_z/\Delta_0<1.055$ is mapped to almost the whole sector V plus sector I). The reason is that near $\Delta_z/\Delta_0=1.045$, the integral on the r.h.s. of Eq. \ref{eq:finaldeltaself} (besides the factor $U_{\alpha\beta}$) for $U_{22}$ passes 0, but the integral for $U_{11}$ remains finite. This says that the order parameter is not sensitive to $U_{22}$, so the same state spans a large domain in the $U$-space.

Moreover, as $m_0$ goes to $-m_0$, the two orbitals are effectively interchanged, so the phase diagrams are mirror image of each other by the $45^\circ$ diagonal line $U_{22}=U_{11}$. This symmetry has been addressed previously in S5.1. Below we show the phase diagrams for the cases of $m_0=\pm3,\pm1$ and $0.1$ explicitly. The $m_0=3$ diagram is explained in detail, while the other diagrams only have quantitative difference so can be understood easily.

\subsection{Phase diagrams of $m_0=\pm3$}
\begin{figure}[h]
\captionsetup{singlelinecheck = false, justification=raggedright}
\centering
\begin{subfigure}{0.32\textwidth}
	\centering
	\includegraphics[width=0.9\textwidth]{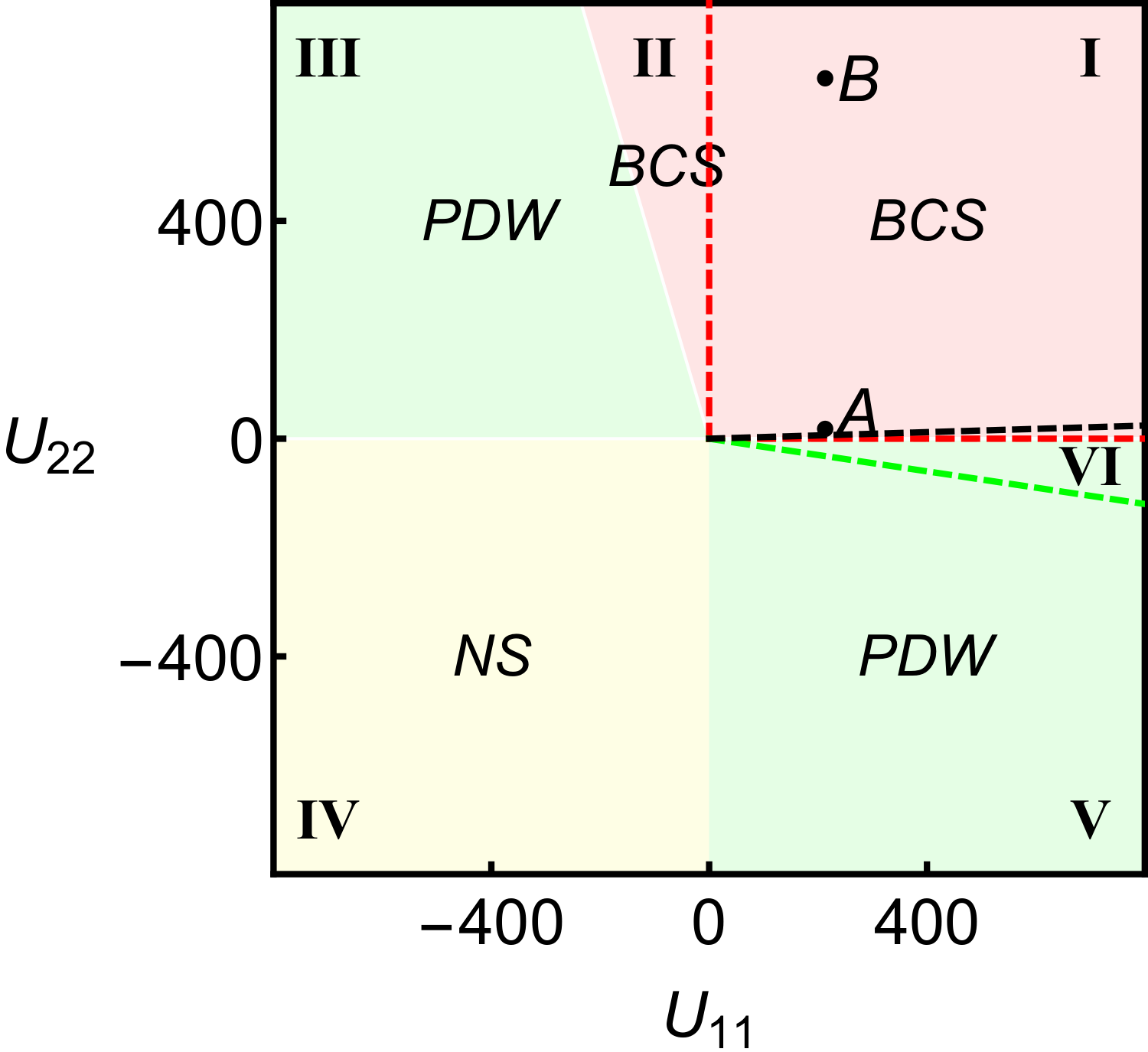}
	\caption{\centering}
\end{subfigure}
\begin{subfigure}{0.32\textwidth}
	\centering
	\includegraphics[width=0.9\textwidth]{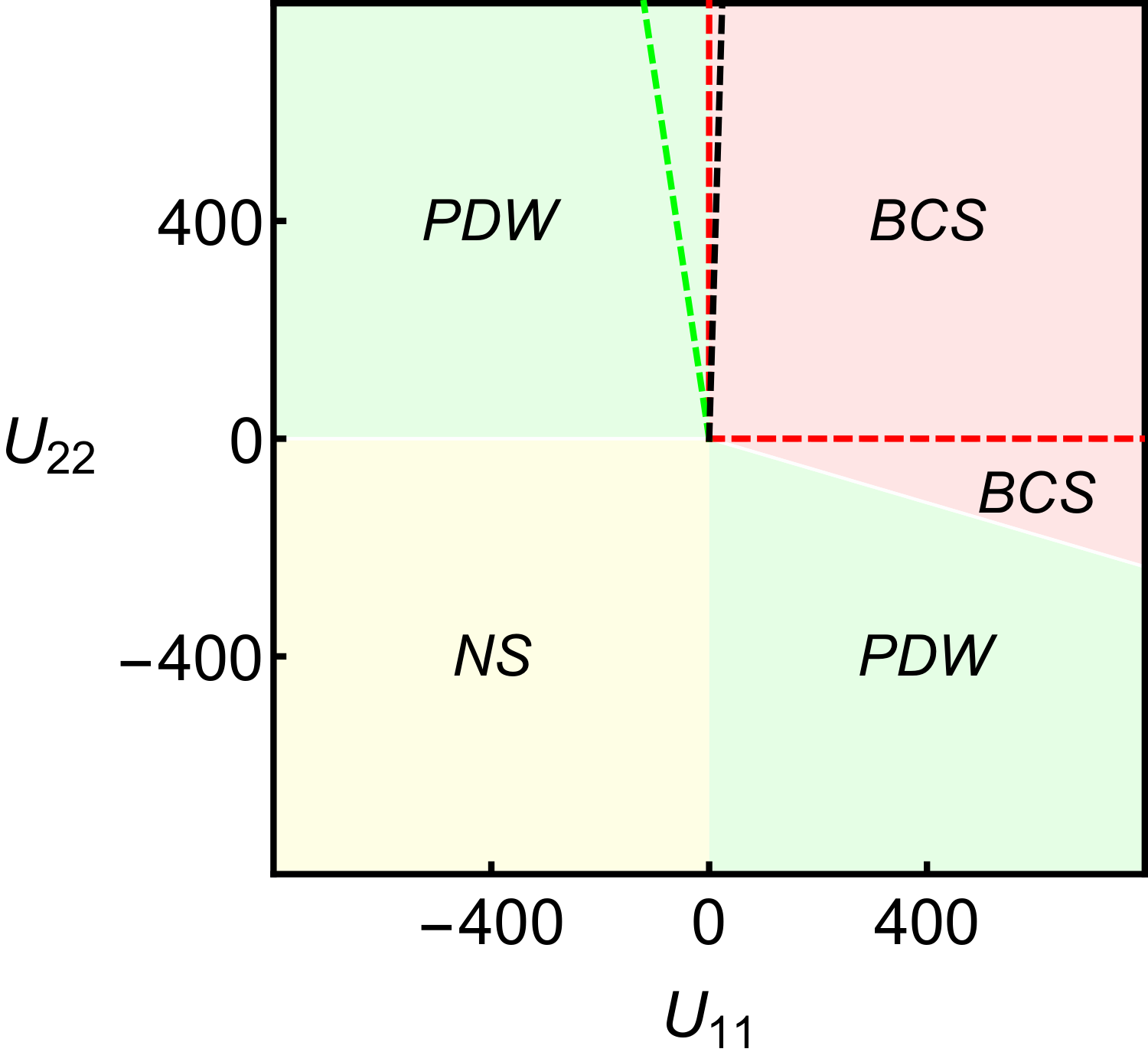}
	\caption{\centering}
\end{subfigure}
	\caption{BCS-PDW phase diagram in the $U_{11}-U_{22}$ space of flattened BHZ model for (a) $m_0=3$ and (b) $m_0=-3$, at nearly half filling. $U_{\alpha\beta}$ are in arbitrary unit and $U_{\alpha\beta}>0$ means attractive interaction.  $\Delta_z/\Delta_0=0$ is mapped to the black dashed line. $\Delta_z/\Delta_0=\pm1$ is mapped to the horizontal and vertical red dashed lines, respectively, which are boundaries of the first quadrant. The green dashed line in (a) and (b) correspond to value $\Delta_z/\Delta_0\approx 1.054$ and $-1.054$, respectively. Diagram (b) is the mirror image of (a) about diagonal line $U_{22}=U_{11}$.}
\label{fig:udiagram1}
\end{figure}

In Fig. \ref{fig:udiagram1} (a), $m_0=3$, we first start from an identity order parameter $\Delta_z/\Delta_0=0$ and move towards the $\Delta_z/\Delta_0>0$ side. Due to the strong polarization to orbital 2, $\Delta_z/\Delta_0=0$ corresponds to the black dashed line, with $U_{22}/U_{11}\approx 0$, which is a BCS state. As we increase $\Delta_z/\Delta_0$, we rotate clockwise in the $U$-space. At $\Delta_z/\Delta_0=1$, $\hat{\Delta}_{22}=0$ so $U_{22}=0$, which corresponds to the horizontal red dashed line. Next, we enter the fourth quadrant by keeping increasing $\Delta_z/\Delta_0$. Right after reaching $\Delta_z/\Delta_0=1$, four nodal circles show up (see Fig. \ref{fig:nodalcircle}(a)) so the BCS-PDW transition occurs immediately and as a result the whole fourth quadrant is a PDW phase.

Special attention should be paid to the value $\Delta_z/\Delta_0\approx 1.054$ and $\Delta_z/\Delta_0\approx 1.11$. $\Delta_z/\Delta_0\approx 1.054$ starts from the green dashed line in Fig. \ref{fig:udiagram1} (a), when the integral in Eq. \ref{eq:finaldeltaself} for $U_{22}$ vanishes and passes 0, making the order parameter insensitive to $U_{22}$. Therefore from $\Delta_z/\Delta_0=1.054$ to $1.055$, the phase point quickly rotates from the green dashed line to negative $U_{22}$ axis, suddenly switching to positive $U_{22}$ axis (the vertical red dashed line), returning to the first quadrant, and then quickly rotates clockwise to the horizontal red dashed line again. This means sector V and sector I always have a PDW solution (channel), corresponding to almost the same ratio $\Delta_z/\Delta_0=1.054\sim 1.055$.

The first quadrant assumes a BCS solution too. The BCS channel is strong and always wins out against the PDW channel, which will be discussed shortly, therefore the whole first quadrant is a BCS phase. At $1.055<\Delta_z/\Delta_0<1.11$, the state stays in the first quadrant at $U_{22}/U_{11}\approx0$. At $\Delta_z/\Delta_0=1.11$, the integral in Eq. \ref{eq:finaldeltaself} for $U_{11}$ passes 0, so it switches from positive $U_{11}$ axis to negative $U_{11}$ axis, keeping rotating clockwise but maintaining $U_{22}/U_{11}\approx0$ until $\Delta_z/\Delta_0=+\infty$. It will connect to the PDW state on the $\Delta_z/\Delta_0<0$ side there.

The $\Delta_z/\Delta_0<0$ side has a much simpler story. In Fig. 3(a) of the maintext, we only showed the $\Delta_z/\Delta_0>0$ side, but the $\Delta_z/\Delta_0<0$ side is centrosymmetric to it, according to the discussions in S5.1. Starting from the black dashed line and rotating counterclockwise, we enter the second quadrant at $\Delta_z/\Delta_0=-1$. The $m_0=3$ phase on the $\Delta_z/\Delta_0<0$ side has no nodal circles (same as $m_0=-3$ phase on the $\Delta_z/\Delta_0>0$ side), so the BCS-PDW transition does not occur immediately and the BCS phase extends into the second quadrant (sector II, also see the leftmost panel of Fig. 3(a) in the maintext). The BCS channel in sector II gets no competition from a PDW channel.

\subsection{Phase diagrams of $m_0=\pm1$ and $0.1$}
\begin{figure}[h]
\captionsetup{singlelinecheck = false, justification=raggedright}
\centering
\begin{subfigure}{0.32\textwidth}
	\centering
	\includegraphics[width=0.9\textwidth]{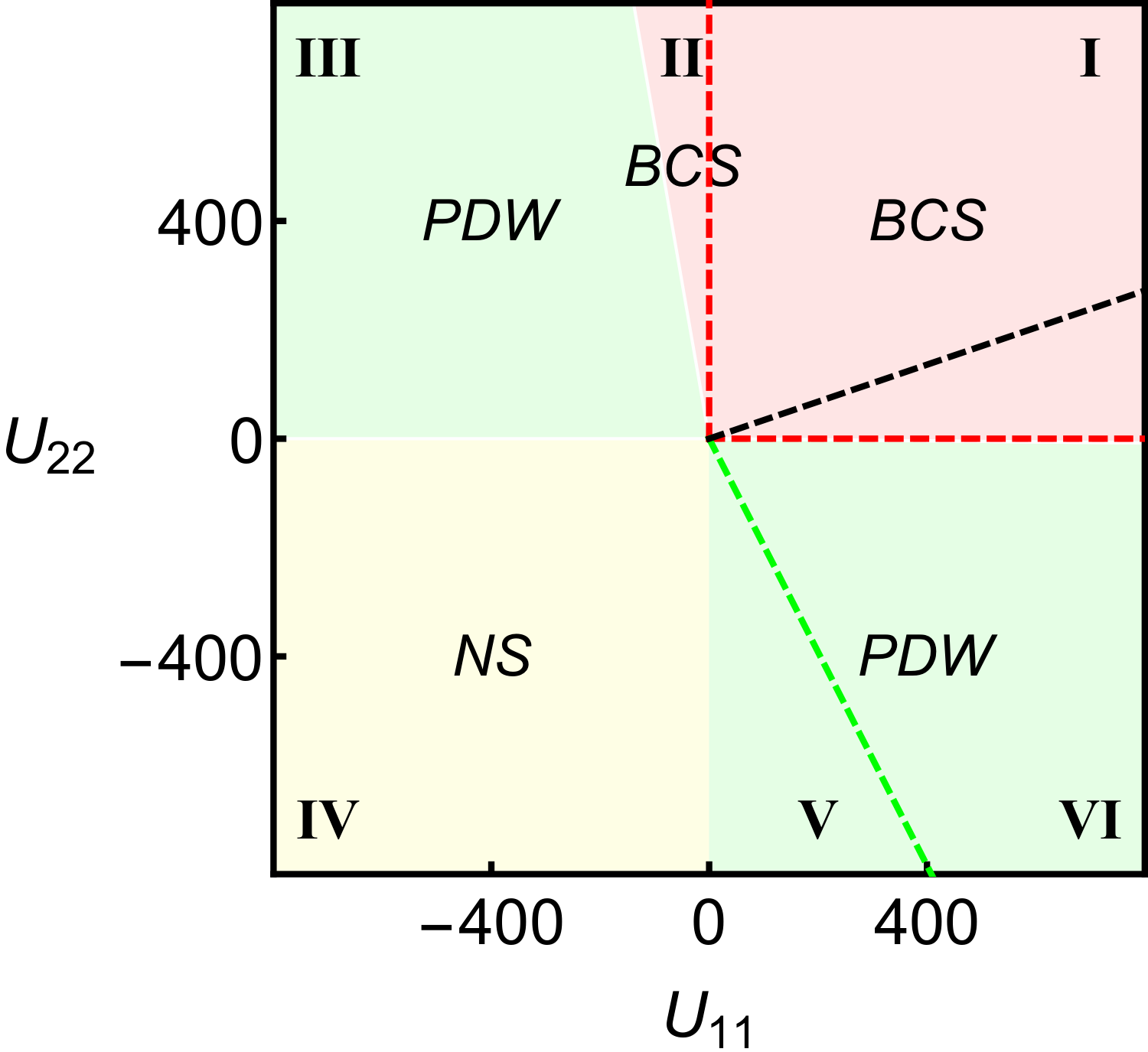}
	\caption{\centering}
\end{subfigure}
\begin{subfigure}{0.32\textwidth}
	\centering
	\includegraphics[width=0.9\textwidth]{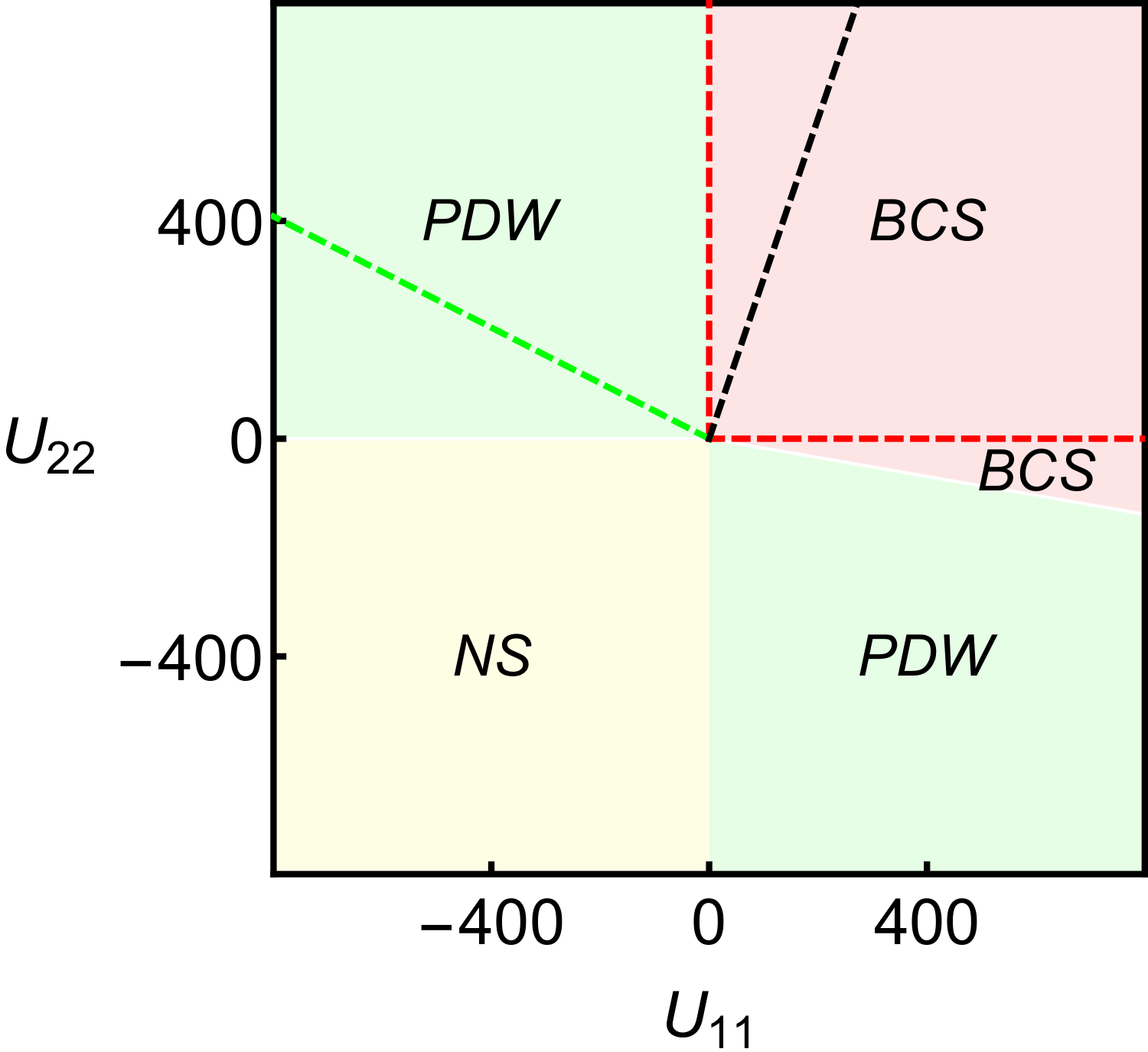}
	\caption{\centering}
\end{subfigure}
\begin{subfigure}{0.32\textwidth}
	\centering
	\includegraphics[width=0.9\textwidth]{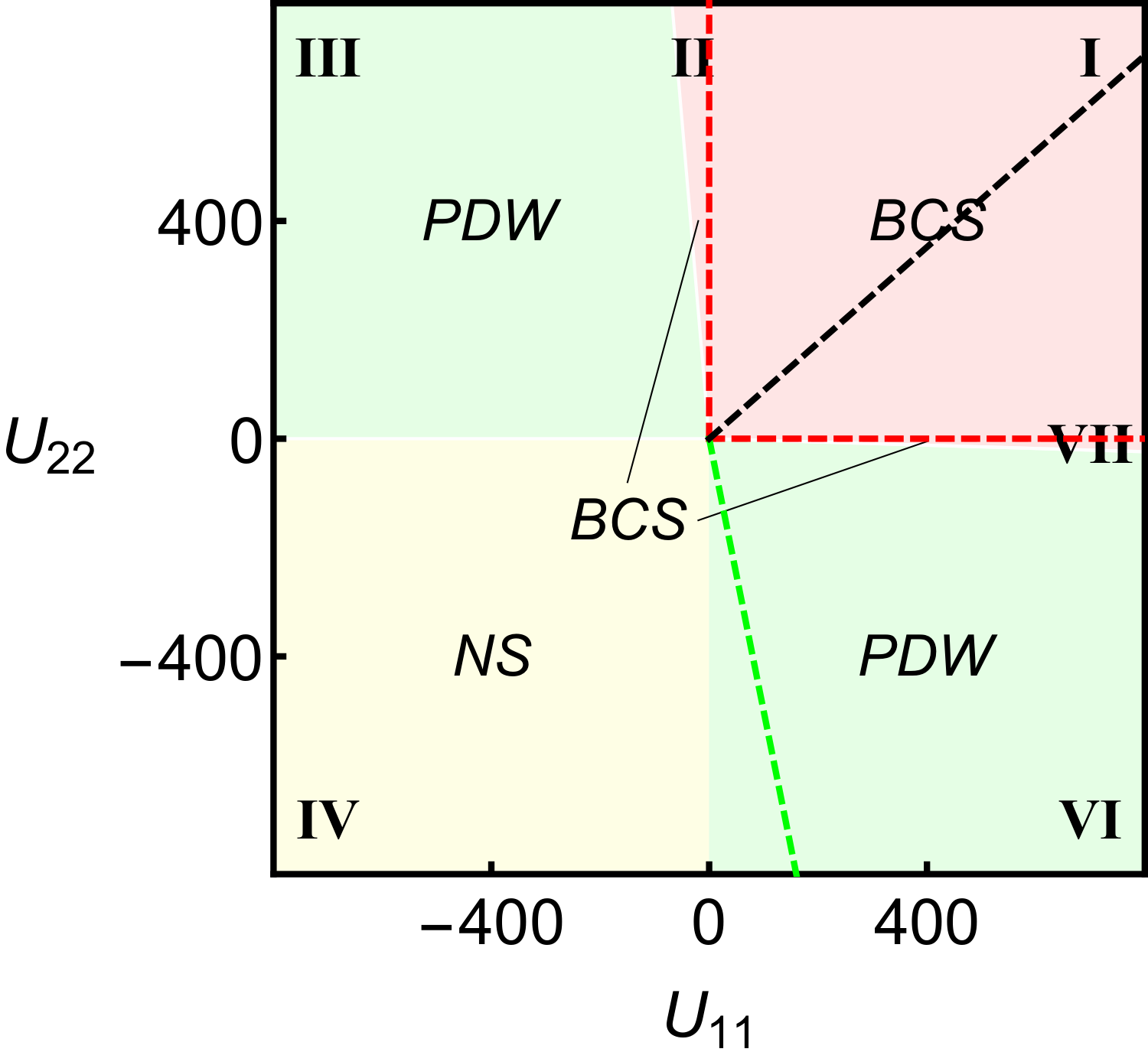}
	\caption{\centering}
\end{subfigure}
\caption{BCS-PDW phase diagram in the $U_{11}-U_{22}$ space of flattened BHZ model for for (a) $m_0=1$, (b) $m_0=-1$ and (c) $m_0=0.1$, at nearly half filling. $U_{\alpha\beta}$ are in arbitrary unit and $U_{\alpha\beta}>0$ means attractive interaction. Here the green dashed line in (a) and (b) correspond to value $\Delta_z/\Delta_0\approx 1.22$ and $-1.22$, respectively; in (c), due to the loss of polarization, the green dashed line gets very close to boundary of 3rd and 4th quadrant. Diagram (b) is the mirror image of (a) about diagonal line $U_{22}=U_{11}$.}
\label{fig:udiagram2}
\end{figure}

For the phase $m_0=1$ (Fig. \ref{fig:udiagram2}(a)), the orbital polarization is not so strong as $m_0=3$. The black dashed line gets closer to $45^\circ$ diagonal line and the difference between $\Delta_z/\Delta_0>0$ and $<0$ side are not so big as $m_0=3$. On the $\Delta_z/\Delta_0>0$ side there are 3 nodal circles, one of which has a large circumference (Fig. \ref{fig:nodalcircle}(b)), making the BCS-PDW transition on entering the second quadrant. In contrast, the $\Delta_z/\Delta_0<0$ side has 1 nodal circle, but with very small circumference when entering the second quadrant (Fig. \ref{fig:nodalcircle}(c)). This is not strong enough to make an immediate BCS-PDW transition, so a small sector II still remains.

For the phase $m_0=0.1$ (Fig. \ref{fig:udiagram2}(b)), the case of little orbital polarization, sector II keeps shrinking and a new small BCS domain in the fourth quadrant (sector VII) starts to develop, making the diagram more symmetric.

\subsection{Competition between BCS and PDW channels in the first quadrant}
\begin{figure}[h]
\captionsetup{singlelinecheck = false, justification=raggedright}
\centering
\begin{subfigure}{0.48\textwidth}
	\centering
	\includegraphics[width=0.8\textwidth]{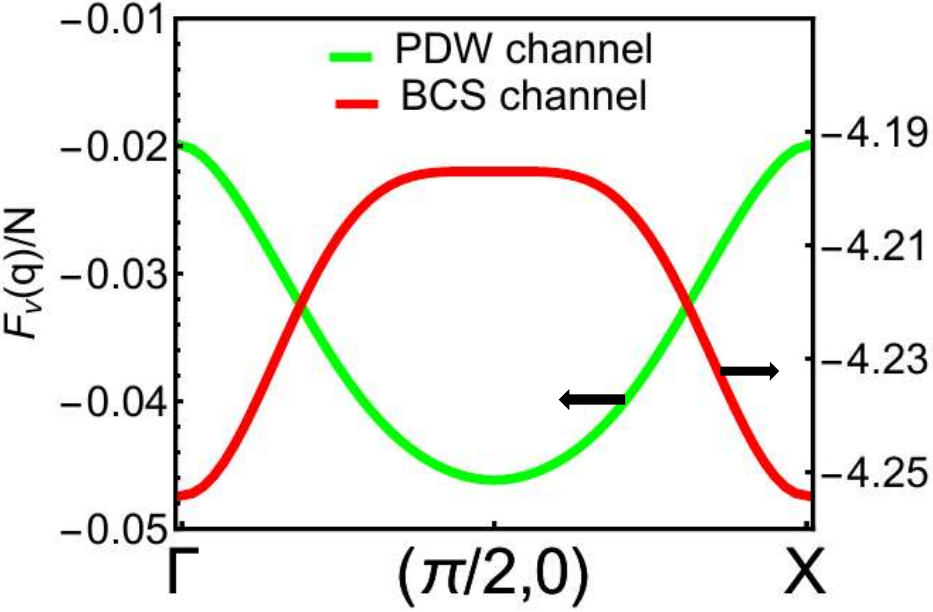}
	\caption{\centering}
\end{subfigure}
\begin{subfigure}{0.48\textwidth}
	\centering
	\includegraphics[width=0.835\textwidth]{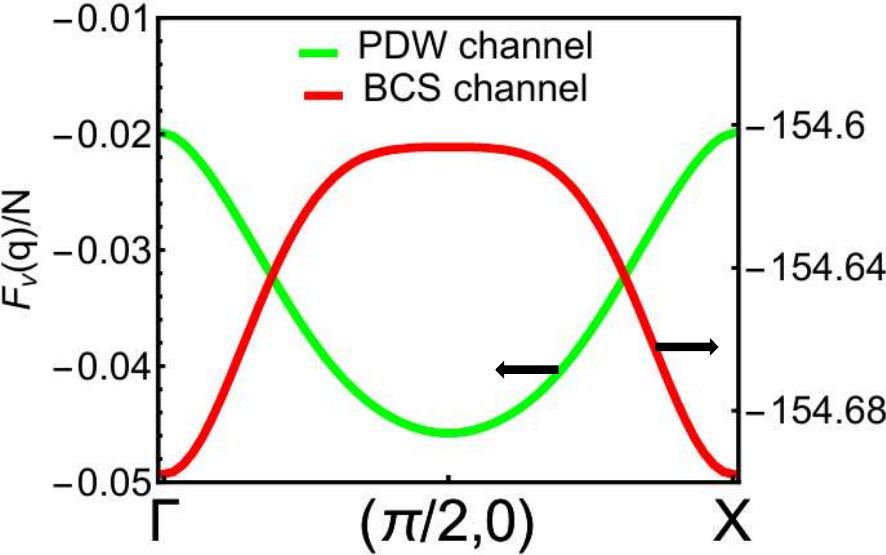}
	\caption{\centering}
\end{subfigure}
	\caption{Competition of BCS and PDW channels in the first quadrant in $U$-space, for Sample point A, B from Fig. \ref{fig:udiagram1}(a). (a) A point, with $U_{11}=213.4$, $U_{22}=18.02$, $n_{e,v}=1.067$. The two solutions are $\Delta_0=1,\Delta_z=1.055,\mu=0.001$ ($\Delta_z/\Delta_0=1.055$ so is PDW) and $\Delta_0=6.08,\Delta_z=-2.62,\mu=0.57$ ($\Delta_z/\Delta_0=-0.43$ so is BCS). (b) B point, with $U_{11}=211.6$, $U_{22}=663.8$, $n_{e,v}=1.067$. The two solutions are $\Delta_0=1,\Delta_z=1.0545,\mu=0.001$ ($\Delta_z/\Delta_0=1.0545$ so is PDW) and $\Delta_0=161.9,\Delta_z=-158.5,\mu=20.8$ ($\Delta_z/\Delta_0=-0.98$ so is BCS).}
\label{fig:twochannel}
\end{figure}

Finally, we discuss the competition between the BCS and PDW channels in the first quadrant (sector I). We call it a BCS or PDW channel if the free energy $F_v(\bd{q})$ of the channel has absolute minima at $\bd{q}=0$ or $\bd{q}\neq0$. Throughout the first quadrant of $U$-space, there is always a BCS solution and a PDW solution, which are both extended $s$-waves, and the BCS channel always win. For example, in Fig. \ref{fig:twochannel} we give the plot of free energy along $\Gamma X$ in $\bd{q}$ space for sampling point A, B from Fig. \ref{fig:udiagram1}(a). To compare the two channels, we must fix the electron density to be the same. The BCS channel turns out to have an order parameter of much larger scale, so is stabler than the PDW channel.

To close this section, we summarize a few points from the $U$-space phase diagram calculations.
\begin{enumerate}
\item Robust PDW phases are observed in the second and fourth quadrant universally for all topological/trivial BHZ phases, which is when the interaction is attractive on one orbital and repulsive on the other.
\item The third quadrant has no nonzero solution, so is a normal state; the second and fourth quadrant has a unique PDW solution; due to the singular behavior of self-consistency equations, the first quadrant is double-valued, with the BCS channel always stabler than the PDW channel, so is a BCS phase.
\item Different $m_0$ values have different strength of orbital polarization, making the diagrams quantitatively different.
\item When there is no nodal circles or when the nodal circle is weak due to broadening, the BCS phase will extends into the second and fourth quadrant, reducing the area of PDW phases.
\end{enumerate}

The last point 4 supports the central idea of this paper---we need either relatively flat bands or in the circumstances of dispersive bands require nodal zeroes for the geometric BCS-PDW transitions.


\begin{thebibliography}{54}
\bibitem{footnote1}
The superfluid weight $D_{s,\mu \nu}$ can be non-positive definite 
for particle-hole asymmetric bands, as in the case of exciton condensates. 
However, for superconductors, the system obeys particle-hole symmetry.
\bibitem{footnote2}
  Here, we only consider superconducting states. The possibilities of 
  other competing symmetry-broken states are beyond the 
  scope of this work.
\bibitem{footnote3}
With our convention $\bd{q}$ is half of the center-of-mass
 momenta of the Cooper pairs in contrast with some literature.
\bibitem{footnote4}
This choice of a hermitian pairing matrix
  $\hat{\Delta}$ corresponds to a special type of spinful time-reversal
  symmetry, which has the operator form $\hat{\mathcal{I}}_s\otimes(-i\hat{s}_y K)$, with
  $\hat{\mathcal{I}}_s$ the identity matrix on the $s$-dim orbital space, $\hat{s}_y$ the {Pauli}
  matrix on spins, and $K$ the complex conjugate operator.
\end{thebibliography}
\end{document}